\documentclass[twocolumn]{aastex63}

\usepackage{amsmath}
\usepackage{amssymb}
\usepackage{color}
\usepackage{url}
\usepackage{ulem}
\usepackage{multirow}
\usepackage{graphics,graphicx}
\usepackage{blindtext}
\usepackage{hyperref}
\usepackage{longtable}
\usepackage{array}
\usepackage{multirow}
\usepackage{savesym}
\savesymbol{tablenum}
\usepackage{siunitx}
\restoresymbol{SIX}{tablenum}

\usepackage{enumerate}
\usepackage{lineno}

\newcommand{\teff}{$T_{\text{eff}}$\xspace}%
\newcommand{\fkep}{$F_{\text{Kep}}$\xspace}%

\newcommand{\gaia}{\textit{Gaia}\xspace}%
\newcommand{\npbar}{$\bar{N}_p$\xspace}

\newcommand{\tdd}{\textit{TD}/\textit{D}\xspace}%
\newcommand{\tdh}{\textit{TD}/\textit{H}\xspace}%
\newcommand{\dher}{\textit{D}/\textit{Her}\xspace}%
\newcommand{\tdher}{\textit{TD}/\textit{Her}\xspace}%
\newcommand{\RNum}[1]{\uppercase\expandafter{\romannumeral #1\relax}}

\begin{document}

\title{Planets Across Space and Time (PAST) \RNum{4}: The Occurrence and Architecture of Kepler Planetary Systems as a Function of Kinematic Age Revealed by the LAMOST-Gaia-Kepler Sample}

\correspondingauthor{Ji-Wei Xie}
\email{jwxie@nju.edu.cn}


\author{Jia-Yi Yang}
\affiliation{School of Astronomy and Space Science, Nanjing University, Nanjing 210023, China}
\affiliation{Key Laboratory of Modern Astronomy and Astrophysics, Ministry of Education, Nanjing 210023, China}

\author{Di-Chang Chen}
\affiliation{School of Astronomy and Space Science, Nanjing University, Nanjing 210023, China}
\affiliation{Key Laboratory of Modern Astronomy and Astrophysics, Ministry of Education, Nanjing 210023, China}

\author{Ji-Wei Xie}
\affiliation{School of Astronomy and Space Science, Nanjing University, Nanjing 210023, China}
\affiliation{Key Laboratory of Modern Astronomy and Astrophysics, Ministry of Education, Nanjing 210023, China}

\author{Ji-Lin Zhou}
\affiliation{School of Astronomy and Space Science, Nanjing University, Nanjing 210023, China}
\affiliation{Key Laboratory of Modern Astronomy and Astrophysics, Ministry of Education, Nanjing 210023, China}

\author{Subo Dong}
\affiliation{Kavli Institute for Astronomy and Astrophysics, Peking University, Beijing 100871, People’s Republic of China}

\author{Zi Zhu}
\affiliation{School of Astronomy and Space Science, Nanjing University, Nanjing 210023, China}
\affiliation{Key Laboratory of Modern Astronomy and Astrophysics, Ministry of Education, Nanjing 210023, China}

\author{Zheng Zheng}
\affiliation{ Department of Physics and Astronomy, University of Utah, Salt Lake City, UT 84112, USA}

\author{Chao Liu}
\affiliation{National Astronomical Observatories, Chinese Academy of Sciences, Beijing 100012, People’s Republic of China}

\author{Weikai Zong}
\affiliation{Department of Astronomy, Beijing Normal University, Beijing 100875, China}
\affiliation{Institute for Frontiers in Astronomy and Astrophysics, Beijing Normal University, Beijing 102206, China}

\author{Ali Luo}
\affiliation{National Astronomical Observatories, Chinese Academy of Sciences, Beijing 100012, People’s Republic of China}

\begin{abstract}
One of the fundamental questions in astronomy is how planetary systems form and evolve.
Measuring the planetary occurrence and architecture as a function of time directly addresses this question.
In the fourth paper of the Planets Across Space and Time (PAST) series, we investigate the occurrence and architecture of Kepler planetary systems as a function of kinematic age by using the LAMOST-Gaia-Kepler sample.
To isolate the age effect, other stellar properties (e.g., metallicity) have been controlled.
We find the following results.
(1) The fraction of stars with Kepler-like planets (\fkep) is about 50\% for all stars; no significant trend is found between \fkep and age.
(2) The average planet multiplicity (\npbar) exhibits a decreasing trend ( $\sim$2$\sigma$ significance) with age.
It decreases from \npbar$\sim$ 3 for stars younger than 1 Gyr to \npbar  $\sim$1.8 for stars about 8 Gyr.
(3) The number of planets per star ($\eta$=\fkep$\times$\npbar) also shows a decreasing trend ($\sim$2--3$\sigma$ significance).
It decreases from $\eta\sim$ 1.6--1.7 for young stars to $\eta\sim$ 1.0 for old stars.
(4) The mutual orbital inclination of the planets ($\sigma_{i,k}$) increases from $1\fdg2^{+1.4}_{-0.5}$ to $3\fdg5^{+8.1}_{-2.3}$ as stars aging from 0.5 to 8 Gyr with a best fit of $\log{\sigma_{i,k}}=0.2+0.4\times\log{\frac{\text{Age}}{\text{1Gyr}}}$.
Interestingly, the Solar System also fits such a trend.
The nearly independence of \fkep$\sim50\%$ on age implies that planet formation is robust and stable across the Galaxy history.
The age dependence of \npbar and $\sigma_{i,k}$ demonstrates planetary architecture is evolving, and planetary systems generally become dynamically hotter with fewer planets as they age.

\end{abstract}


\section{Introduction}
Thanks to various surveys from ground \citep[e.g.,][]{Mayor.2011arXiv1109.2497M} and space \citep[e.g.,][]{Borucki_2010_Sci_327_977B}, the number of known planets has reached a milestone \citep[5000, NASA Exoplanet Archive\footnote{\url{https://exoplanetarchive.ipac.caltech.edu/}},][]{Akeson.2013PASP..125..989A}.
Such a rich planetary database has enabled substantial statistical studies of the occurrence rate and architecture of planetary systems \citep[see the reviews by][]{Winn_2015_ARA&A_53_409W,Zhu.2021ARA&A..59..291Z}, deepening our understanding of planet formation and evolution. 

Stellar properties (e.g., mass, effective temperature, and metallicity) play crucial roles in determining planetary occurrence rate and architecture.
Although the occurrence of giant planets (Jupiter-like gas giants) has been found to increase with stellar mass \citep{Johnson_2010_PASP_122_905J,Ghezzi.2018ApJ...860..109G}, the trend is opposite for small planets.
For the bulk of planets detected by the Kepler mission (so called super-Earths and sub-Neptunes, with radii between the Earth and Neptune, hereafter dubbed as Kepler planets for short), their occurrence rate in terms of number of planets per star ($\eta$) has an inverse relationship with stellar temperature and mass \citep{Howard_2012_ApJS_201_15H,Mulders_2015_ApJ_798_112M,Kunimoto.2020AJ....159..248K}.
In fact, $\eta$ can be further decomposed into two factors: the fraction of stars that have planetary systems ($F$) and the average number of planets in a planetary system (planetary multiplicity $\bar{N}_p$), and they are linked by the following equation:
\begin{equation}{\label{eq:eta}}
    \eta=F\times\bar{N}_p.
\end{equation}
Further studies have shown that both $F$ and \npbar tend to decrease as stellar temperature and mass increase \citep{Yang.2020AJ....159..164Y,He.2021AJ....161...16H}.

Metallicity also plays a differential role in shaping planetary systems of giant planets and small planets.
On one hand, a correlation between giant planets and metallicity has been well established \citep{Santos.2001A&A...373.1019S,Fischer.2005ApJ...622.1102F}, which provides the key evidence for the core-accretion model of planet formation \citep[e.g.,][]{Pollack.1996Icar..124...62P,Ida_2004_ApJ_604_388I}.
On the other hand, such a planet-metallicity correlation is generally weaker and more complicated for small planets \citep{Buchhave.2012Natur.486..375B,Wang.2015AJ....149...14W,Dong.2018PNAS..115..266D,Petigura.2018AJ....155...89P,Zhu.2019ApJ...873....8Z}

Stellar environments (e.g., stellar companions, clusters and memberships of the Galactic thin/thick disks) also affect planetary occurrence and architecture. 
There has been substantial evidence showing that planetary occurrence is reduced and planetary architecture is modified when stellar companions are close, with separations $\lesssim$ 100 AU \citep{Wang.2014ApJ...783....4W,Kraus.2016AJ....152....8K,Moe.2021MNRAS.507.3593M,Fontanive.2019MNRAS.485.4967F,Su.2021AJ....162..272S}.
Recently, it has been reported that both the period and radius distributions of exoplanets exhibit dependencies on stellar clustering  \citep{Winter.2020Natur.586..528W,Kruijssen.2020ApJ...905L..18K,Chevance.2021ApJ...910L..19C,Longmore.2021ApJ...911L..16L}.
\citet{Dai.2021AJ....162...46D} found that stellar groups with high relative velocities tend to have a lower occurrence rate of super-Earths and sub-Neptunes but a higher occurrence rate of sub-Earths.
The Galactic membership and total velocity of stars are also linked with the planet occurrence rate.
It has been found that stars in the thick disk (higher total velocity) generally have fewer planets than those in the thin disk \citep[lower total velocity,][]{Bashi.2019AJ....158...61B,Bashi.2022MNRAS.510.3449B,Chen.2021AJ....162..100C}.

The occurrence and architecture of planets in our Galaxy could also evolve with time.
Therefore, measuring planet occurrence and architecture as a function of time can provide crucial insights into planet formation and evolution.
For example, recent studies \citep[e.g.,][]{Berger_radius_2020AJ....160..108B,Sandoval.2021ApJ...911..117S,David.2021AJ....161..265D,Chen.2022AJ....163..249C} have revealed that the relative occurrence (ratio) of super-Earths and sub-Neptunes evolves on a time scale of Giga years, providing crucial constraints on the formation of the radius valley \citep[a deficit of planets with radii of $\sim$1.7-2.1 $R_{\oplus}$,][]{Fulton.2017AJ....154..109F}.
More recently, \citet{Bashi.2022MNRAS.510.3449B} found tentative evidence that suggests the occurrence rate of close in super-Earths detected by Kepler is anti-correlated with stellar age. 
However, such an anti-correlation is still inconclusive, probably because they adopted the isochrone ages which suffer from large uncertainties \citep[56\% for Kepler stars,][]{Berger.2020AJ....159..280B}. 
Furthermore, they didn't isolate the effect of age from other stellar properties (e.g., metallicity), so it is still unclear whether the anti-correlation is intrinsic or just a projection of other correlations.  

To investigate how planet occurrence and architecture evolve with time, we have started a series of work named Planets Across Space and Time \citep[PAST,][]{Chen.2021ApJ...909..115C}.
The first challenge of this work is to determine the age of main-sequence stars, which make up the bulk of planet hosts.
In PAST \RNum{1} \citep{Chen.2021ApJ...909..115C}, we revisited the kinematic method to classify Galactic components and the Age-Velocity dispersion Relationship \citep[AVR,][]{Stromberg.1946ApJ...104...12S,Wielen.1977A&A....60..263W,Holmberg.2009A&A...501..941H}, extending the viable range to 1.5 kpc to cover the majority of the known exoplanet hosts.
The deduced kinematic age for an ensemble of stars has a typical internal uncertainty of 10\%--20\%.
The second challenge is to isolate the effect of stellar age, because age is generally correlated with other properties, such as stellar mass and metallicity.
Applying the revised kinematic method of PAST \RNum{1}, we constructed a catalog of kinematic and other basic properties for 35,835 Kepler stars in PAST \RNum{2} \citep{Chen.2021AJ....162..100C}.
Such a large and homogeneous sample enables us to further set control samples to isolate the effect of age from other stellar properties.
In PAST \RNum{3}, we investigated how the radius distribution of small planets evolves with time \citep{Chen.2022AJ....163..249C}.
In this work, the fourth paper of the PAST series (PAST \RNum{4}), we study the occurrence and architecture of Kepler planets as a function of stellar age.

This paper is organized as follows: in Section \ref{sec:data}, we present the star and planet data used in this work. In Section \ref{sec:pc}, we describe the parameter control method to isolate the age effect, and present the apparent occurrence of Kepler planets. In Section \ref{sec:result}, we adopt a forward modeling method to derive the intrinsic occurrence rate and architecture of Kepler planets. We make the discussions and summarize the main conclusions in Section \ref{sec:discussion} and \ref{sec:conclusion}.

\section{Data sample}{\label{sec:data}}
\subsection{Star Sample}
The LAMOST-Gaia-Kepler catalog constructed in PAST \RNum{2} is based on LAMOST DR4/DR5 and GAIA DR2.
Since LAMOST have updated to DR8 \citep{Yan.2022Innov...300224Y}, and Gaia has released EDR3/DR3 \citep{Gaia.2022arXiv220800211G}, therefore we have updated the LAMOST-Gaia-Kepler catalog accordingly.
We start from the stellar properties catalogue from \citet{Berger.2020AJ....159..280B}, which provides a homogeneous calibration of effective temperature, mass, and radius for most of the Kepler stars.
The Kepler team calculated Combined Differential Photometric Precision \citep[CDPP,][]{Christiansen.2012PASP..124.1279C,keplerstar,keplercomplete} for each target, which defines the completeness of transit searching.
We restrict our sample to targets with $\sigma_{\text{CDPP}}$ value \footnote{\url{https://exoplanetarchive.ipac.caltech.edu/docs/Kepler_completeness_reliability.html}}.
Then we cross-match the sample with the recently released LAMOST DR8 \footnote{\url{http://www.lamost.org/dr8/}} low-resolution catalogue \citep{Yan.2022Innov...300224Y}, which contains more spectral observations of Kepler stars reprocessed with the latest pipeline version.
We select stars with LAMOST metallicity and radial velocity measurements, and remove stars with [Fe/H] less than $-1.0$ due to a lack of training set \citep{Xiang.2019ApJS..245...34X}.
Next, we cross-match with Gaia DR3\citep{Gaia.2022arXiv220800211G,gaiadr3} catalog, which includes more accurate measurements of positions, proper motions, and parallaxes of stars compared to DR2.
Gaia DR3 also provides the renormalised unit weight error \citep[RUWE,][]{LL:LL-124} for identifying possible binary stars.
Stars with RUWE values greater than 1.2 are excluded from our sample \citep{Berger.2020AJ....159..280B,Bryson.2020AJ....159..279B}.
We obtain 70,239 star in total, the number of our sample after each selection step is summarized in Table \ref{table:data}.

\begin{figure}[htb!]
\centering
\includegraphics[width=0.45\textwidth]{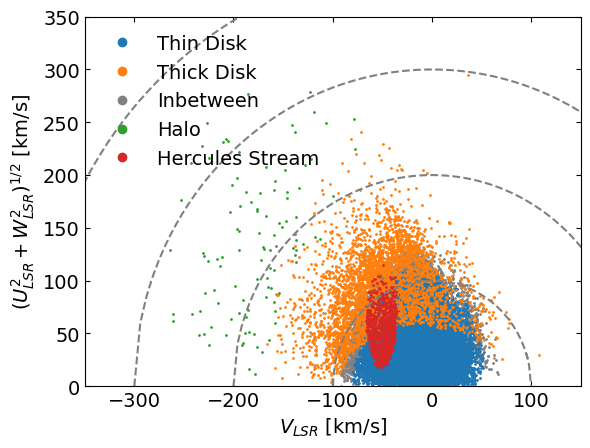}
\caption{Toomre diagram of stars in the updated LAMOST-Gaia-Kepler sample. The blue, orange, grey, green, red, and grey dots present stars in the thin disk, thick disk, in between thin and thick disk, halo, and Hercules stream, respectively. The grey dot lines represent the total Galactic velocity at 100, 200, and 300 $\unit{km.s^{-1}}$.
\label{fig:fig_toomre}}
\end{figure}

Utilizing Gaia DR3, we update the kinematic method and AVR of PAST \RNum{1}, the details can be seen in the Appendix \ref{sec:app_edr3}.
In PAST \RNum{1}, the calibrations of the kinematic method and AVR extended from the solar neighborhood to a larger range of stars with $|Z|<1.5$ kpc, $7.5<R<10$ kpc, and distance$<$1.5 kpc, where $Z$ and $R$ are the vertical and radial components of Galactocentric cylindrical coordinates, respectively.
Here, we adopt a similar range of stars, but further extend the distance to 2 kpc, thanks to the improvement of the astrometric measurements from Gaia DR2 to Gaia DR3.
With the updated kinematic method (see Appendix \ref{sec:app_edr3}), we calculate the probabilities of stars belonging to each Galactic component, i.e, thin disk, thick disk, halo, and Hercules stream (dub as $D$, $TD$, $H$, and $Her$).
We classify stars into different components following the commonly used method introduced by \citet{Bensby.2003A&A...410..527B,Bensby.2014A&A...562A..71B}, and show the results in the Toomre diagram (Figure \ref{fig:fig_toomre}).
Since AVR can only be applied to disk stars, so we limit our sample to stars within the Galactic disk, i.e., stars with $D/Her\geqslant2$, $TD/H\geqslant1$, and $TD/Her\geqslant2$.

\begin{figure}[htb!]
\centering
\includegraphics[width=.45\textwidth]{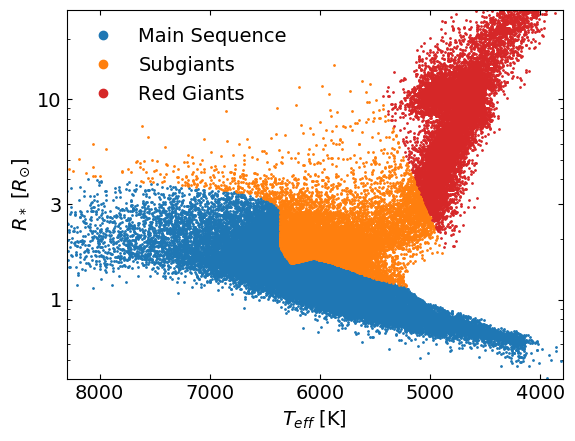}
\caption{Hertzsprung--Russell diagram of stars in the updated LAMOST-Gaia-Kepler sample. The blue, orange, and red dots show stars in the main-sequence stage, subgiant, and red giant, respectively.
\label{fig:fig_hr}}
\end{figure}

In Figure \ref{fig:fig_hr}, we show stars from the updated LAMOST-Gaia-Kepler sample in the Hertzsprung--Russell diagram.
The effective temperature and radius data are obtained from \citet{Berger.2020AJ....159..280B}, and the evolve stage is calculated by the same method as \citet{Bryson.2020AJ....159..279B} using the python package \texttt{evolstate}\footnote{\url{http://ascl.net/1905.003}}.
We further limit our star sample to main-sequence solar type stars, with effective temperature between 4700 to 6500 $\unit{\kelvin}$.
So far, the number of our star sample is 19,537.

\subsection{Planet Sample}
\begin{figure}[htb!]
\centering
\includegraphics[width=.45\textwidth]{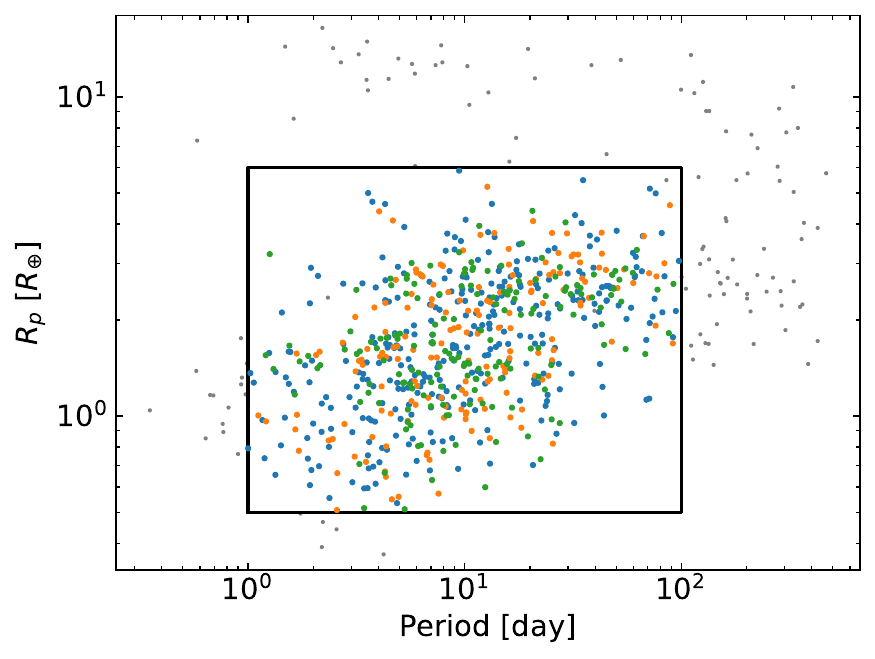}
\caption{Planet sample in the period-radius diagram. The grey dots show the whole planet sample before data selection. The blue, orange, and green dots present planets in single, double, and three or more planet systems after we apply all the selection criteria. 
\label{fig:fig_pla}}
\end{figure}

{\label{sec:data_pla}}
Our Kepler planet sample is based on Kepler DR25 \citep{Thompson.2018ApJS..235...38T,koidr25}.
We select planets/candidates within our star sample, and exclude those flagged as `false positives'.
We only consider planets with periods less than 100 days, as the observed planet numbers and detection efficiencies both drop significantly beyond this period \citep{Burke.2017ksci.rept...19B}.
We also exclude planetary systems with Ultra Short Period planets (USPs, period $<$ 1 day) from our Fiducial analysis (see Section \ref{sec:usp} for more discussions on USPs).
This exclusions is due to the standard Kepler pipeline is not well conditioned to search for USPs \citep{Sanchis-Ojeda.2014ApJ...787...47S}, and USPs are relatively rare (with an occurrence rate $\sim$ 1\%) and may have undergone different formation and evolution process \citep{Dai.2018ApJ...864L..38D}.
The planet radii are derived from stellar radii \citep{Berger.2020AJ....159..280B} and the planet-to-star radius ratio \citep[$R_p/R_s$,][]{Thompson.2018ApJS..235...38T}.
Planets with radii smaller than 0.5$R_{\oplus}$ are excluded due to their relatively low detection efficiency \citep{Burke.2017ksci.rept...19B}.
Since we focus on the occurrence rate and architecture of small planets, we exclude planet systems with planets larger than 6$R_{\oplus}$ (see Section \ref{sec:usp} for more discussions on giant planets).
The selection process of the planet sample is summarized in Table \ref{table:data}.
The period-radius distribution of our planet sample is shown in Figure \ref{fig:fig_pla}.

\subsection{Kinematic Age of Planet Host and non-Host}
\begin{figure}[htb!]
\centering
\includegraphics[width=.45\textwidth]{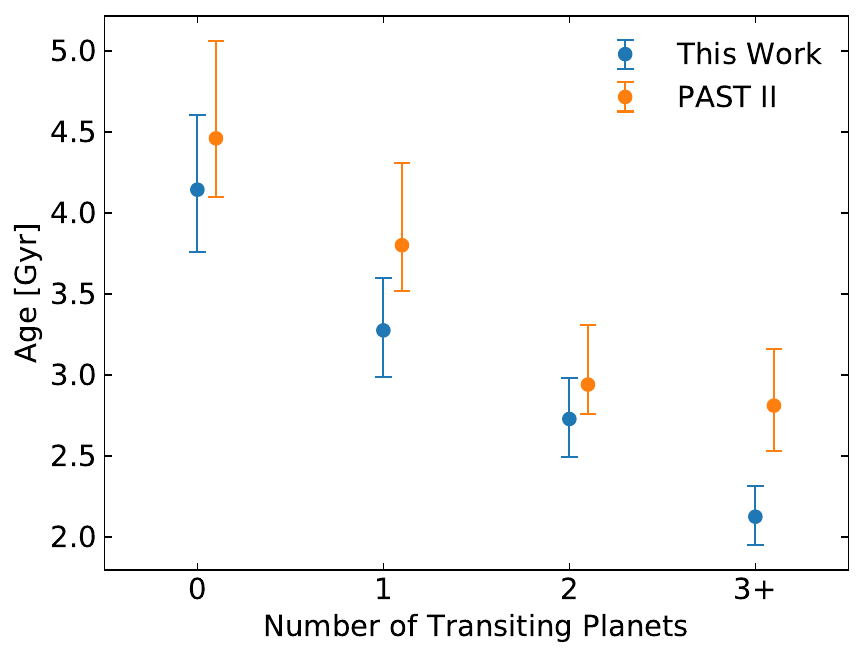}
\caption{The blue dots and errorbars show the kinematic age and $\pm1\sigma$ ranges for stars with zero, one, two, and three or more planets for this work, and the orange dots and errorbars are values from PAST \RNum{2} \citep{Chen.2021AJ....162..100C}, the x-axis is offset by 0.1 for clearance.
\label{fig:fig_age}}
\end{figure}

In PAST \RNum{2}, we have shown that stars with different numbers of transiting planets exhibit a drop in their kinematic age drops as a function of planet multiplicity.
Here, we also bin stars according to their transiting planet number into four groups; in each group, stars have zero, one, two, and three or more transiting planets, respectively.
We derive the kinematic age for each group using the updated AVR (see details in Appendix \ref{sec:app_edr3}).
Since kinematic age is calculated from the dispersion of total Galactic velocity, it can be skewed by velocity outliers.
To reduce the effect caused by outliers, we calculate the median value and the Mean Absolute Deviation (MAD) of the total Galactic velocity for each group.
Then we remove stars with total velocities higher or lower than $\text{Median} \pm5\times$MAD within this group.
The kinematic age results for each group are presented in Figure \ref{fig:fig_age}, and we compare them to the results of PAST \RNum{2} \citep{Chen.2021AJ....162..100C}.
As can be seen, the kinematic ages derived in this work are generally consistent with those of PAST \RNum{2} in $1\sim2\sigma$ range.
They both show a declining trend in kinematic age with increasing planet multiplicity.
Nevertheless, the ages obtained in this work are systematically lower by about 0.2--0.5 Gyr compared to those in PAST \RNum{2}.
This difference is expected, because we remove outlier stars in this work, which usually have high velocities.
The removal of outliers causes a decrease in velocity dispersion and, consequently, a lower value of kinematic age.

In total, we obtain 19,358 stars in the the star sample, and 663 planets in 467 systems.
The size of our sample after each step of selection can be seen in Table \ref{table:data}.

\startlongtable
\begin{deluxetable}{lrr}
\tablewidth{0pt}
\tablecaption{Data selection
\label{table:data} }
\tablehead{
\colhead{ } &\colhead{Star}   &\colhead{Planet}}
\startdata 
	\citet{Berger.2020AJ....159..280B}      &186,301    &3,826\\
	With $\sigma_{\text{CDPP}}$                           &185,161    &3,826\\
	Match with LAMOST DR8                   &70,251     &1,562\\
	With RV data and $\text{[Fe/H]}\geqslant-1$     &68,567     &1,549\\
    Math with \gaia DR3                    &67,922     &1,535\\
    RUWE$\leqslant$1.2 (Remove potential binary) &55,332     &1,320\\
	\hline
	$|Z|<$1.5 kpc, 7.5$<R<$10 kpc, & &\\
	and distance$<$2 kpc          &45,914     &1,279\\
	\tdh$\geqslant$1, \tdher$\geqslant$2, \dher$\geqslant$2   &40,347     &1,109\\
	(In the thin, thick disk, or in between) \\
	\hline
	Main sequence                           &27,213     &940\\
	4700$\unit{\kelvin}\leqslant$\teff$\leqslant6500\unit{\kelvin}$       &19,537     &784\\
	\hline
	Orbit Period$\leqslant$100 days         &...        &720\\
	Remove ultra short period system        &...        &703\\
	(No planet with period $<$1 day)\\
	Planet radii$\geqslant$0.5 RE           &...        &698\\
	Remove giant planet system              &...        &663\\
	(No planet with radii $>$6 RE)\\
	\hline
	Median$-5\times$MAD$\leqslant V_{tot}$   &  &  \\
	and $V_{tot}\leqslant$Median$+5\times$MAD  &19,358     &641\\
\enddata
\end{deluxetable}

\section{Apparent trend analysis from parameter control}{\label{sec:pc}}
In this section, we derive the apparent planet occurrence rate as a function of stellar kinematic age.
The apparent planet occurrence rate is defined as the \emph{observed} planet multiplicity function (number of stars that have one, two, and three or more transit planets, dubbed as $N_1$, $N_2$, and $N_{3+}$) divided by the number of stars ($N_{star}$) in each bin.

\subsection{Parameter Control}{\label{sec:pc_method}}

To isolate the effect caused by stellar age, we use the parameter control method to reduce the influences induced by other stellar properties.
In this work, we control five properties: effective temperature, mass, metallicity, stellar radius, and $\sigma_{\text{CDPP}}$.
The former three parameters need to be controlled because they are found to affect the intrinsic planet occurrence rate \citep[e.g.,][]{Buchhave.2012Natur.486..375B,Yang.2020AJ....159..164Y,He.2021AJ....161...16H}.
The latter two also need to be controlled because they directly affect the detection efficiency of transiting planets.

The basic idea of the parameter control is to let stars in different age bins have similar distributions in the controlled stellar properties.
To achieve this goal, we apply a `finding star neighbors' method, similar to the one described in \citet{Chen.2022AJ....163..249C}, which involves the following steps:

\begin{enumerate}[{(}1{)}]
\item Grouping stars. For the whole star sample with a size of $N$, we sort the stars according to \tdd ascendingly, which is an effective indicator of stellar age \citep{Chen.2021ApJ...909..115C}.
Then we group the stars into an odd number of bins.
To implement parameter control, the middle bin contains fewer stars, while the other bins have more stars. 
The farther away from the middle bin, the more stars there are.
In this study, we first consider a case of three bins, each bin containing 40\%, 20\%, and 40\% of the stars, to have a qualitative view of the age trend.
To further quantify the age--occurrence rate trend, we consider a case of five bins, each containing 25\%, 20\%, 10\%, 20\%, and 25\% of the stars, respectively.
Due to the limited sample size, we do not consider cases with more bins.

\item \label{itm:standard} Choosing a standard sample. We dub the stars in the middle bin as the `standard sample', and the number of stars in the central bin is denoted as $N_{st}$.

\item \label{itm:nearest} Finding the nearest neighbor stars. 
In each bin (except the middle one), we select $N_{st}$ stars that are the closest neighbors of the standard sample in the space of the controlled parameters. 
This is done by adopting the nearest neighborhood method from the \texttt{scikit-learn} \citep{Pedregosa.2012arXiv1201.0490P} package.

\item \label{itm:percentile} Checking parameter control result. 
We calculate the differences of the 25, 50, and 75 percentiles of each controlled parameter for every two bins.
If all the differences are less than the typical errors, we consider these parameters have been controlled.
The typical errors of temperature, mass, and radius are 112\unit{\kelvin}, 7\%, and 4\%, respectively \citep{Berger.2020AJ....159..280B}.
For metallicity we choose 0.05 dex as the typical error, which is the median value of the internal measurement uncertainties in our star sample.
The Kepler team has reported $\sigma_{\text{CDPP}}$ for different timescales.
We choose $\sigma_{\text{CDPP}}$ of 3.5 hours because it is the closest to the median transit duration of our planet sample (3.35 hours).
Since the SNR of Kepler planet is in proportion to $(R_p/R_s)^2/\sigma_{\text{CDPP}}$, we choose a typical error of 10\% for $\sigma_{\text{CDPP}}$, to match the uncertainty induced by stellar radius and planet-star radius ratio.

\end{enumerate}

\begin{figure}[htb!]
\centering
\includegraphics[width=.5\textwidth]{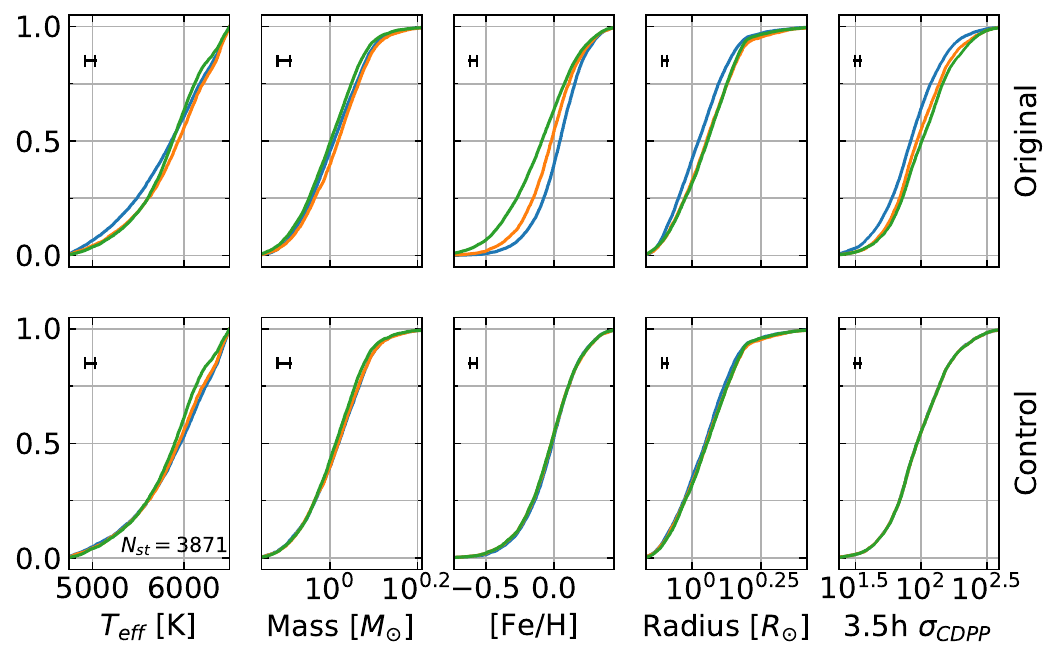}
\caption{Cumulative Distribution Function (CDF) diagrams of effective temperature, mass, metallicity, radius, and $\sigma_{\text{CDPP}}$ 3.5 hours for the three bins method before and after parameter control. 
The errorbar in the upper left corner shows the typical error of each stellar property, and the number in the lower right corner presents the size of the standard sample.
\label{fig:bin3_cdf}}
\end{figure}

\begin{figure}[htb!]
\centering
\includegraphics[width=.5\textwidth]{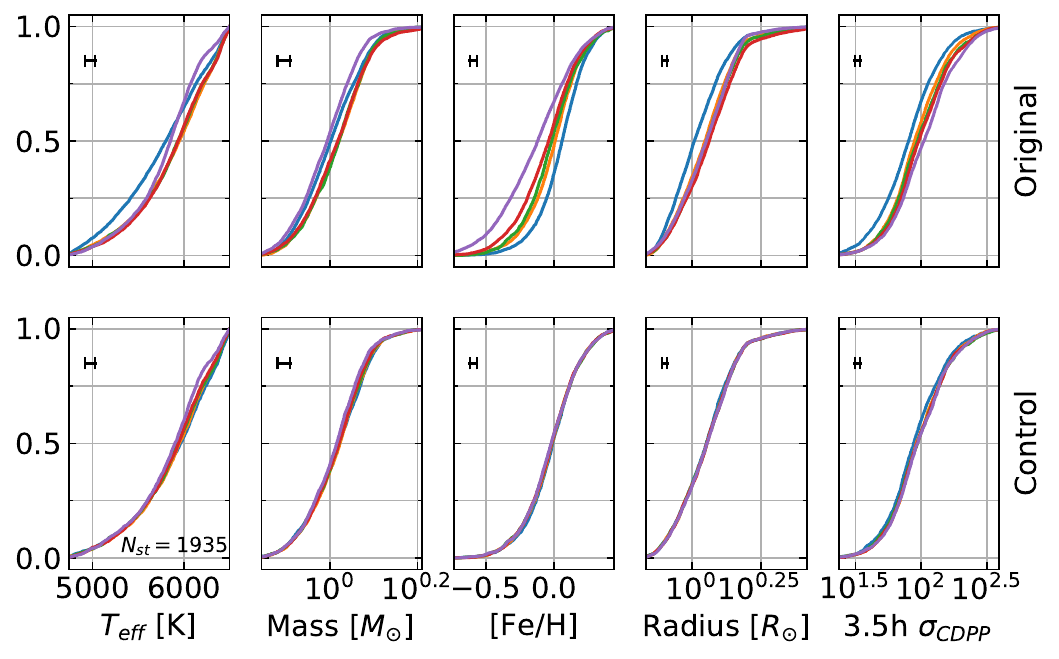}
\caption{Similar to Figure \ref{fig:bin3_cdf}, here we show CDF diagrams for the five bins case. The blue, orange, green, red, and purple lines represent star data sample with \tdd in the ranges of (0,0.0427), [0.0427--0.0618), [0.0618-0.0.0808), [0.0808--0.184], and $[0.184,+\inf$).
\label{fig:bin5_cdf}}
\end{figure}

To get an intuitive view of how well the parameters have been controlled, we plot Figure \ref{fig:bin3_cdf} and Figure \ref{fig:bin5_cdf} in which we perform parameter control for the cases of three age bins and five age bins.
In the first row of Figure \ref{fig:bin3_cdf} and Figure \ref{fig:bin5_cdf}, we show the Cumulative Distribution Function (CDF) diagrams of temperature, mass, metallicity, radius, and $\sigma_{\text{CDPP}}$ 3.5 hours for the observation data.
Using the above method, we control all parameters and show the CDF of controlled star samples in the bottom row.
By applying the parameter control method, we have achieved the goal to let stars in different age bins have similar distribution in stellar temperature, mass, metallicity, radius, and $\sigma_{\text{CDPP}}$ (Figure \ref{fig:bin3_cdf} and \ref{fig:bin5_cdf}).

\subsection{Apparent Planetary Occurrence as a Function of Age} 
\begin{figure}[thb!]
\centering
\includegraphics[width=.45\textwidth]{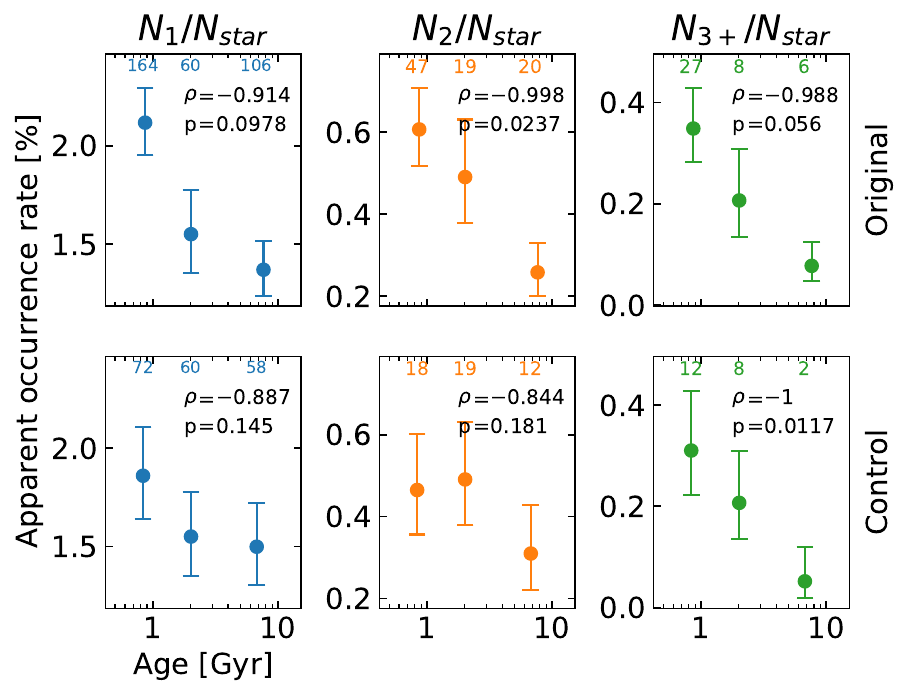}
\caption{Apparent planet occurrence rate for the three bins case with one (left), two (middle), and three or more (right) transiting planets. 
The top and bottom rows correspond to the results before and after parameter control, as shown in Figure \ref{fig:bin3_cdf}. 
The dots and errorbars present the median value and $\pm1\sigma$ range, assuming Poisson error.
The numbers at the top of each panel show the corresponding planet system numbers ($N_1$, $N_2$, and $N_{3+}$).
In the corner of each panel, we also print the Pearson correlation coefficient ($\rho$) and the corresponding $p$-value.
\label{fig:bin3_fs}}
\end{figure}

{\label{sec:pc_results}}
We first consider a three bins case and calculate the kinematic age of each bin using AVR (Appendix \ref{sec:app_edr3}).
We adopt the above controlled sample and calculate the numbers of stars that have one, two, three or more planets, i.e., the planet multiplicity function ($N_1$, $N_2$, and $N_{3+}$).
The apparent occurrence rate of one, two, and three or more planet systems is derived by dividing $N_1$, $N_2$, and $N_{3+}$ by the number of stars ($N_{star}$) in each bin.
In Figure \ref{fig:bin3_fs}, from the left column to the right column, we present the apparent occurrence rate for one, two, and three or more planet systems.
As can be seen, the young stars generally have a higher apparent occurrence than the old stars.
For the original data without parameter control (first row of Figure \ref{fig:bin3_fs}), $N_1/N_{star}$, $N_2/N_{star}$, and $N_{3+}/N_{star}$ are $2.12^{+0.18}_{-0.17}$\%, $0.61^{+0.10}_{-0.09}$\%, and $0.35^{+0.08}_{-0.07}$\% for stars less than 1 Gyr, which are about 4.8$\sigma$, 4.4$\sigma$, and 4.8$\sigma$ higher than those ($1.37^{+0.15}_{-0.13}$\%, $0.26^{+0.07}_{-0.06}$\%, and $0.08^{+0.05}_{-0.03}$\%) for stars about 8 Gyr, respectively.
For the data after all parameter control (bottom row of Figure \ref{fig:bin3_fs}), $N_1/N_{star}$, $N_2/N_{star}$, and $N_{3+}/N_{star}$ are $1.86^{+0.25}_{-0.22}$\%, $0.46^{+0.14}_{-0.11}$\%, and $0.31^{+0.12}_{-0.09}$\% for the youngest group, which are about 1.6$\sigma$, 1.4$\sigma$, and 3.3$\sigma$ higher than those ($1.50^{+0.22}_{-0.20}$\%, $0.31^{+0.12}_{-0.09}$\%, and $0.05^{+0.07}_{-0.03}$\%) for the oldest group, respectively.
The differences in apparent rate between young and old groups become smaller when taking into account of parameter control.
Such a change is more prominent for low multiplicity systems ($N_1/N_{star}$ and $N_2/N_{star}$) than for high multiplicity systems ($N_{3+}/N_{star}$).
We also calculate the Pearson correlation coefficients and $p$-values for the correlations between age and apparent planet occurrence, which are printed in each panel of Figure \ref{fig:bin3_fs}.
The $p$-values are derived using the following steps.
\begin{enumerate}[{(}1{)}]
    \item We calculate the Pearson correlation coefficients for the observation data as $\rho_{obs}$.
    \item \label{itm:scramble} We generate simulated apparent occurrence rates for each bin assuming Poisson error, and randomly scramble their order.
    Then we calculate the Pearson correlation coefficient between age and the scrambled data ($\rho_{sim}$).
    \item We repeat Step \ref{itm:scramble} for 10,000 times and calculate the fraction of the simulated data that produce a stronger anti-correlation, i.e., $\rho_{sim}<\rho_{obs}$. 
    This fraction gives the $p$-value of the Pearson correlation for the observed data.
\end{enumerate}
As we can see, the anti-correlations between age and $N_1/N_{star}$, $N_2/N_{star}$ become weaker after parameter control, with the $p$-values rising from 0.0978 to 0.145 and from 0.0237 to 0.181, respectively.
Nevertheless, the anti-correlation between age and $N_{3+}/N_{star}$ becomes stronger, with the $p$-value decreasing from 0.056 to 0.0117.

\begin{figure}[tb!]
\centering
\includegraphics[width=.45\textwidth]{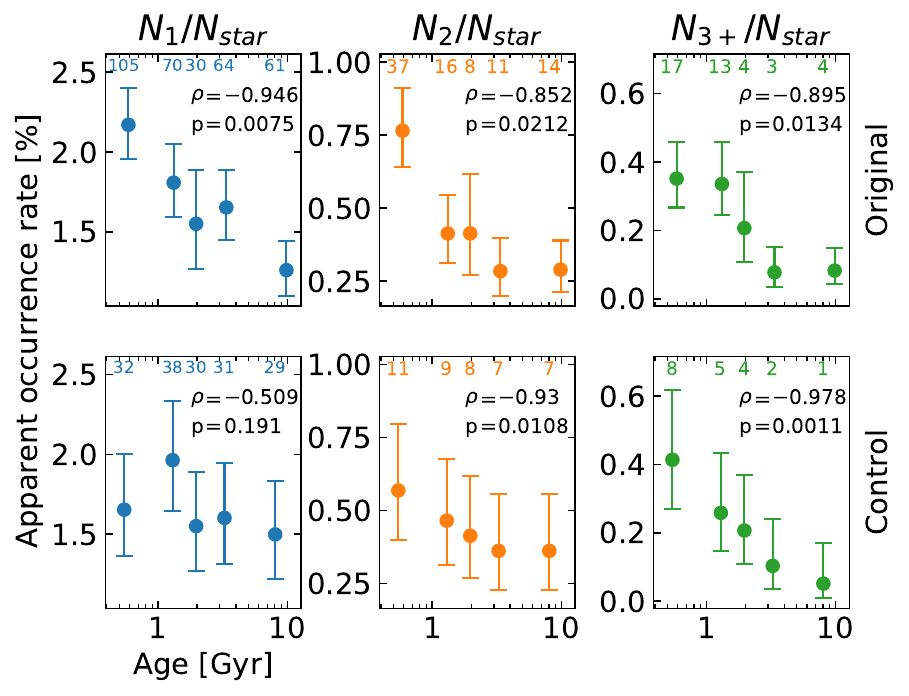}
\caption{Apparent planet occurrence as a function of kinematic age for systems with one (left), two (middle), and three or more (right) transiting planets. 
The top and bottom rows correspond to the results before and after parameter control as shown in Figure \ref{fig:bin5_cdf}.
The dots and errorbars present the median value and $\pm1\sigma$ range assuming Poisson error.
The numbers on the top of each panel show the corresponding planet system numbers ($N_1$, $N_2$, and $N_{3+}$).
In the corner of each panel, we also print the Pearson correlation coefficient ($\rho$) and the corresponding $p$-value.
\label{fig:bin5_fs}}
\end{figure}

To further trace the apparent planet occurrence trend with age, we group the stars into five age bins, and calculate the corresponding kinematic age.
In Figure \ref{fig:bin5_fs}, we present the apparent planet occurrence rate as a function of kinematic age for systems with one, two, and three or more planets.
In line with the results of the three bins case, we find that the apparent occurrence rate generally has a declining trend with kinematic age.
For the original data (before parameter control), $N_1/N_{star}$, $N_2/N_{star}$, and $N_{3+}/N_{star}$ are $2.17^{+0.23}_{-0.21}$\%, $0.76^{+0.15}_{-0.13}$\%, and $0.35^{+0.11}_{-0.08}$\% for the youngest stars, which are 4.6$\sigma$, 4.2$\sigma$, and 3.6$\sigma$ higher than those ($1.26^{+0.18}_{-0.16}$\%, $0.29^{+0.10}_{-0.08}$\%, and $0.08^{+0.07}_{-0.04}$\%) for the oldest stars, respectively (top row of Figure \ref{fig:bin5_fs}).
For the data after all parameter control, $N_1/N_{star}$, $N_2/N_{star}$, and $N_{3+}/N_{star}$ are $1.65^{+0.35}_{-0.29}$\%, $0.57^{+0.23}_{-0.17}$\%, and $0.41^{+0.20}_{-0.14}$\% for stars less than 1 Gyr, which are about 0.5$\sigma$, 1.1$\sigma$, and 2.8$\sigma$ higher than those ($1.50^{+0.33}_{-0.28}$\%, $0.36^{+0.19}_{-0.13}$\%\%, and $0.05^{+0.12}_{-0.04}$\%) for stars about 8 Gyr, respectively (bottom row of Figure \ref{fig:bin5_fs}).
Again, being consistent with the results in the three bins case, the differences in the apparent occurrence rate of stars with different ages become smaller after parameter control.
Nevertheless, the differences are still significant ($\sim3\sigma$) for systems with high transiting multiplicities ($N_{3+}/N_{star}$, right column of Figure \ref{fig:bin5_fs}).
We also calculate the Pearson correlation coefficients and $p$-values, as in the three bins case, and print them in each panel of Figure \ref{fig:bin5_fs}.
Similar to the three bins case, the anti-correlation between age and $N_1/N_{star}$ becomes weaker after parameter control, with the $p$-value rising from 0.0075 to 0.191.
Nevertheless, the anti-correlations between age and multiple planet systems ($N_2/N_{star}$ and $N_{3+}/N_{star}$) become a little stronger, with the $p$-values dropping from 0.212 to 0.0108 and from 0.0134 to 0.0011, respectively.

We have also employed Canonical Correlation Analysis (CCA) to investigate the relationship between stellar properties and the apparent planet occurrence.
The CCA method derives similar results as shown above, indicating that stellar age is anti-correlated with planet occurrence without the need for performing parameter control. 
However, the CCA method is unable to identify the star and planet samples required for the forward modeling method (see Section \ref{sec:result}) in order to derive the intrinsic planet occurrence.  
For more detailed information, please refer to Appendix \ref{sec:cca}.

\section{INTRINSIC trend analysis from forward modeling}\label{sec:result}
\subsection{Forward Modeling Method}
The above apparent occurrence rates only reflect the observed planet population.
In order to derive the intrinsic planet occurrence rates of the underlying planet population, we adopt a forward modeling method that takes into account of \emph{transit} observation bias and detection/vetting efficiencies.
The framework of the method has been described in detail in \citet{Zhu_2018_ApJ_860_101Z} and \citet{Yang.2020AJ....159..164Y}.
In this section, we summarize the general procedure and emphasize the modifications considered in this work.

\subsubsection{General procedure of the modeling}{\label{sec:model_basic}}

We derive the observed planet multiplicity function ($N_1$, $N_2$, and $N_{3+}$) from the star sample.
Then we generate simulated planet systems, and calculate the modeled multiplicity function ($\bar{N}_1$, $\bar{N}_2$, and $\bar{N}_{3+}$).
For our star sample with a size of $\sim20,000$, we need to generate about $\sim10,000$ planet systems, assuming on average 50\% of the stars own planet systems (the true value of \fkep differs for each group and is automatically adjusted in the MCMC process).
The total number of generated planet systems is about 400,000,000 when the simulation is converged.
We assume the multiplicity function follows a Poisson distribution, and optimize the likelihood function
\begin{equation}
	\mathcal{L} = \prod\limits_{k=1}^{3+} \frac{\bar{N}_k^{N_k} \exp (-\bar{N}_k)}{N_k !}
\end{equation}
with \texttt{python} package \texttt{emcee} \citep{Foreman-Mackey.2013PASP..125..306F} applying the Markov Chain Monte Carlo (MCMC) method.
Three free parameters are constrained in our model, which are the fraction of stars with Kepler-like planets (\fkep), the average planet number for stars that have Kepler-like planets (\npbar), and the inclination slope index ($\alpha$, see below).
Details of generating the modelled multiplicity function ($\bar{N}_1$, $\bar{N}_2$, and $\bar{N}_{3+}$) can be seen in \citet{Yang.2020AJ....159..164Y}.
We briefly summarize the general procedure as follows:

\begin{enumerate}[{(}1{)}]
    \item Assuming the intrinsic planet occurrence. 
    For each group of stars, we assume that a fraction of \fkep percent of stars have Kepler-like planets, and on average, each planet system has \npbar planets.
    For each host star, we generate $k$ planets following a zero-truncated Poisson distribution of \npbar \citep{Fang_2012_ApJ_761_92F}.
    
    \item Assuming transit parameters and radii for planets. 
    We generate the debiased distributions of transit parameter ($\epsilon$, $\epsilon=R_s/a_p$, $R_s$ is the stellar radius and $a_p$ is the semi-major axis of the planet) and planet radii ($R_p$) considering three kinds of bias \citep{Mulders.2018AJ....156...24M}, namely, the transit geometry bias ($f_{\text{tra}}$), detection efficiency bias ($f_{\text{S/N}}$), and vetting efficiency bias ($f_{\text{vet}}$).
    For each planet, we assign values of $\epsilon$ and $R_p$ that are randomly drawn from the debiased distributions.
    
    \item Adjusting period ratios and radius ratios. Planets within the same system tend to have similar period ratios \citep{Fabrycky.2014ApJ...790..146F,Brakensiek.2016ApJ...821...47B,Jiang.2020AJ....160..180J} and radius ratios \citep{Ciardi.2013ApJ...763...41C,Weiss.2018AJ....155...48W}.
    To account for this correlation, we adjust the period ratios and radius ratios for planets within the same system. 
    These adjustments are based on debiased distributions calculated by \texttt{CORBITS} \citep{Brakensiek.2016ApJ...821...47B} for multiple planet systems.

    \item Checking Orbital stability. To ensure that the simulated planetary systems are physically plausible, we assess their orbital stability. We apply the criterion provided by \citet{Deck_2013_ApJ_774_129D} to prevent planets within the same system from being located too close to each other.
    
    \item Assigning orbital inclination to generate transits. 
    We calculate the inclination ($I_p$) of the planets with respect to the observer by
    \begin{equation}
        \cos I_p = \cos I \cos i \ -\ \sin I \sin i \cos \phi,
    \end{equation}
    where $I$ represents the inclination of the system invariable plane, $\phi$ is the phase angle, and $i$ is the inclination of the planet with respect to the invariable plane.
    Both $I$ and $\phi$ are assumed to be isotropic.
    Following \citet{Zhu_2018_ApJ_860_101Z}, for a planet system with $k$ planets, the inclination dispersion of the planets follows a power-law function,
    \begin{equation}\label{eq:inclination}
        \sigma_{i,k}\equiv \sqrt{\left<\sin^2i\right>}=\sigma_{i,5}\left(\frac{k}{5}\right)^{\alpha},
    \end{equation}
    where we adopt $\sigma_{i,5}$ as a Gaussian distribution with a mean of $0\fdg8$ and a standard deviation of $0\fdg15$ from \citet{Zhu_2018_ApJ_860_101Z}.
    We fit the inclination slope index $\alpha$ as the third parameter.
    A planet is considered to be a transit if its impact factor is less than 1 ($|\cos{I_p}/\epsilon|<1$).
    
    \item Checking detection and vetting efficiencies. Due to detection and vetting efficiencies, not all the transiting planets can be detected. For each transiting planet, we generate a random number from a uniform distribution ranging from 0 to 1. We consider this planet can be detected if the generated random number is less than the product of the detection efficiency ($f_{\text{S/N}}$, see Appendix \ref{sec:de}) and the vetting ($f_{\text{vet}}$) efficiency.
\end{enumerate}

\subsubsection{Emphasize the difference from \citet{Yang.2020AJ....159..164Y}}{\label{sec:model_diff}}
Comparing to our previous work, we do not consider the TTV multiplicity function, namely, the number of systems that show a TTV signal.
The TTV function is omitted for two reasons.
First, the number of stars in our sample in this work is less than 20,000, and the number of systems that show a TTV signal is only 31.
This is much smaller compared to the 100,000 star sample and 127 systems that show TTV signals in \citet{Yang.2020AJ....159..164Y}.
The smaller number of the TTV multiplicity function leads to larger uncertainty.
Second, as shown in the Appendix of \cite{Yang.2020AJ....159..164Y}, although removing the TTV multiplicity function from the likelihood leads to less constraint on the parameter $\alpha$, it has little impact on the results of \fkep and \npbar, which are the core parameters of this work.

\subsection{Intrinsic Planetary Occurrence and Architecture as a Function of Age}
\begin{figure}[tb!]
\centering
\includegraphics[width=.45\textwidth]{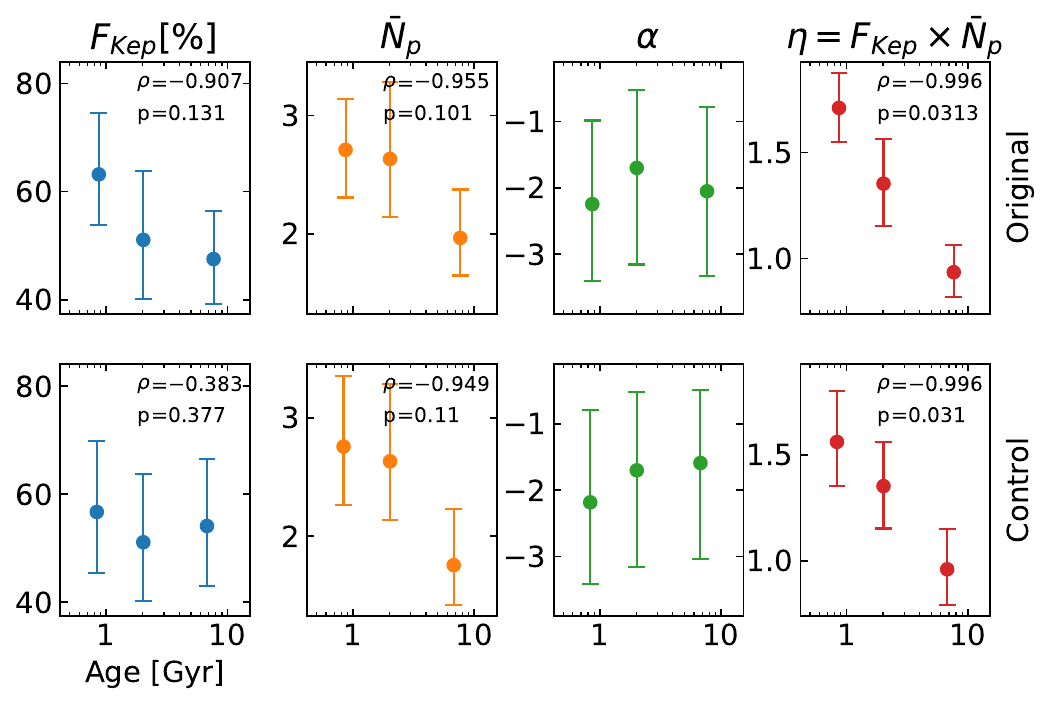}
\caption{Posterior distributions of \fkep, \npbar, $\alpha$, and $\eta$ for the three bins case are presented. The top and bottom rows show the forward modeling results corresponding to samples before and after parameter control as in Figure \ref{fig:bin3_cdf} and Figure \ref{fig:bin3_fs}.
The dots and errorbars show the 50\% and $\pm1$--$\sigma$ range.
In the upper right of the panels in the first, second, and fourth columns, we also print the Pearson correlation coefficient ($\rho$) and the corresponding $p$-value.
\label{fig:bin3_fne}}
\end{figure}

We show the forward modeling results for the case of three bins in Figure \ref{fig:bin3_fne}.
From the left panels to the right panels, we present the posterior distributions of \fkep, \npbar, $\alpha$, and $\eta$ (which is the product of \fkep and \npbar, see Equation \ref{eq:eta}).
For data without parameter control, the youngest group generally has higher intrinsic planet occurrence rates than the oldest group.
\fkep, \npbar, and $\eta$ are $63.2^{+11.4}_{-9.4}$\%, $2.71^{+0.43}_{-0.40}$, and $1.71^{+0.16}_{-0.16}$ for stars less than 1 Gyr, which are 1.7$\sigma$, 1.8$\sigma$, and 5.4$\sigma$ higher than those ($47.5^{+8.9}_{-8.4}$\%, $1.97^{+0.41}_{-0.32}$, and $0.93^{+0.13}_{-0.12}$) for stars about 8 Gyr, respectively (top row of Figure \ref{fig:bin3_fne}).
For the data after all parameter control, \fkep, \npbar, and $\eta$ are $56.9^{+13.3}_{-11.6}$\%, $2.74^{+0.64}_{-0.49}$, and $1.57^{+0.23}_{-0.22}$ for the youngest group, which are 0.2$\sigma$, 2.1$\sigma$, and 3.0$\sigma$ higher than those ($54.3^{+12.3}_{-11.4}$\%, $1.75^{+0.47}_{-0.32}$, and $0.96^{+0.19}_{-0.17}$) for the oldest group, respectively (bottom row of Figure \ref{fig:bin3_fne}).
All the three groups have nearly the same \fkep when parameter control is taken into account.

This result is expected because \fkep is mainly determined by the apparent occurrence rate of single planet systems ($N_1/N_{star}$).
The difference in apparent occurrence rate for single planet systems between the youngest and oldest groups drops from 4.8$\sigma$ to 1.6$\sigma$ after parameter control (Figure \ref{fig:bin3_fs}), and consequently, the difference of \fkep drops from 1.7$\sigma$ to 0.2$\sigma$.

\npbar is largely determined by the apparent occurrence rate of high multiplicity systems ($N_{3+}/N_{star}$).
The difference in $N_{3+}/N_{star}$ between the youngest and oldest groups drops mildly from 4.8$\sigma$ to 3.3$\sigma$ after parameter control (Figure \ref{fig:bin3_fs}).
Interestingly, the difference in \npbar increases slightly from 1.8$\sigma$ to 2.1$\sigma$.
Due to the limited number of planetary systems with three or more planets in the last bin after parameter control (2 systems in the three bins case), the Poisson error is relatively high ($2^{+1.8}_{-1.3}$).
We may have underestimated the declining trend of $N_{3+}/N_{star}$, and the results of the forward modeling show that the decrease in \npbar becomes slightly more prominent.
As $\eta$ is the product of \fkep and \npbar, therefore, the decrease of the difference in $\eta$ from 5.4$\sigma$ to 3.0$\sigma$ is mainly due to the decrease of the difference in \fkep.
We calculate the Pearson correlation coefficients and $p$-values for the correlations between age and the intrinsic planet occurrence (\fkep, \npbar, and $\eta$), and print them in each panel in Figure \ref{fig:bin3_fne}.
The $p$-values are derived using a similar method as shown in Section \ref{sec:pc_results}.
For \fkep and \npbar, the anti-correlations between age and them are statistically insignificant before and after parameter control with $p$-values larger than 0.05.
The anti-correlation between age and $\eta$ maintains with $p$-value smaller than 0.05.

The parameter $\alpha$ is not well constrained in all the cases before and after parameter control, because it is mainly constrained by TTV multiplicity function \citep{Zhu_2018_ApJ_860_101Z}, which is ignored in this work.

\begin{figure}[htb!]
\centering
\includegraphics[width=.45\textwidth]{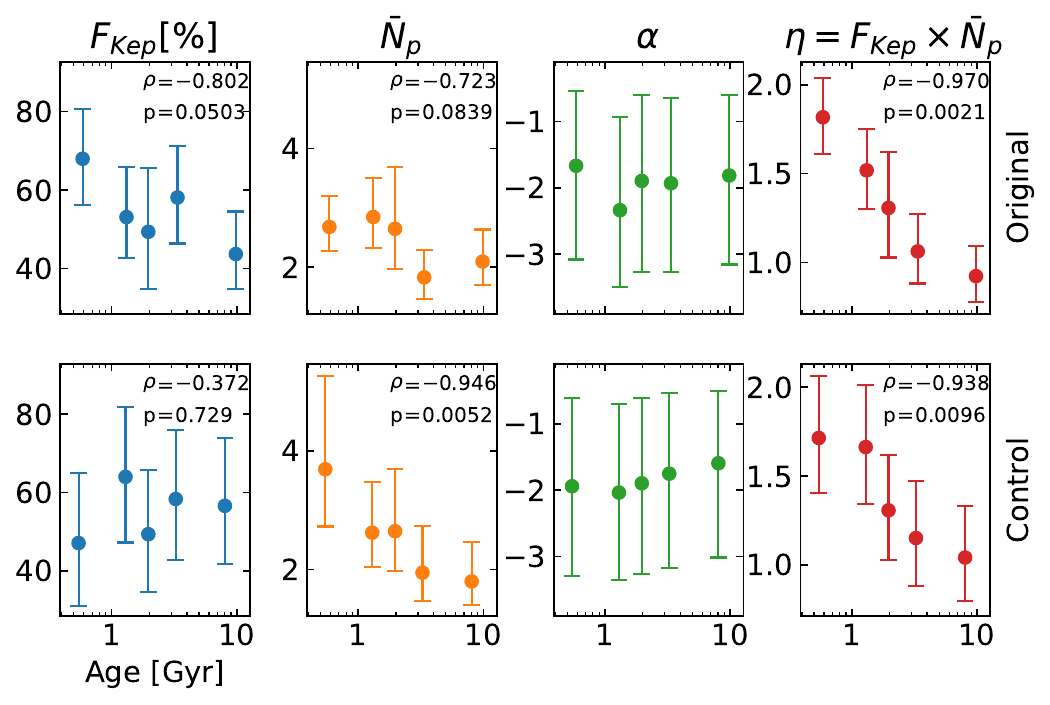}
\caption{Similar to Figure \ref{fig:bin3_fne}, posterior distributions of \fkep, \npbar, $\alpha$, and $\eta$ for the five bins case are presented. The top and bottom rows show the forward modeling results corresponding to samples before and after parameter control, as shown in Figure \ref{fig:bin5_cdf} and Figure \ref{fig:bin5_fs}. The dots and errorbars show the 50\% and $\pm1$--$\sigma$ range.
In the upper right of the panels in the first, second, and fourth columns, we also print the Pearson correlation coefficient ($\rho$) and the corresponding $p$-value.
\label{fig:bin5_fne}}
\end{figure}

To further investigate the intrinsic planet occurrence as a function of age, we group the stars into five bins as mentioned before in Section \ref{sec:pc_results}, and adopt the forward modeling method to derive \fkep, \npbar, $\alpha$, and $\eta$.
The results are shown in Figure \ref{fig:bin5_fne}.
For data without parameter control (top row of Figure \ref{fig:bin5_fne}),  the values of \fkep, \npbar, and $\eta$ are $68.0^{+12.8}_{-11.9}$\%, $2.68^{+0.52}_{-0.41}$, and $1.82^{+0.22}_{-0.20}$, respectively, for stars in the youngest group.
These values are 2.1$\sigma$, 1.2$\sigma$, and 4.8$\sigma$ higher than those ($43.7^{+10.9}_{-8.9}$\%, $2.09^{+0.54}_{-0.39}$, and $0.92^{+0.17}_{-0.15}$) for stars in the oldest group.
After parameter control (bottom row of Figure \ref{fig:bin5_fne}), \fkep is $47.1^{+17.9}_{-16.1}$\% for stars less than 1 Gyr, which is 0.6$\sigma$ lower than stars around 8 Gyr ($56.6^{+17.4}_{-14.9}$\%). 
As to \npbar and $\eta$, they are $3.69^{+1.58}_{-0.96}$ and $1.71^{+0.35}_{-0.31}$ for stars in the first bin, which are 2.3$\sigma$ and 2.2$\sigma$ higher than the values ($1.80^{+0.67}_{-0.39}$ and $1.04^{+0.29}_{-0.24}$) in the last bin.

The forward modeling results for the five bins are consistent with those for the three bins (Figure \ref{fig:bin3_fne}).
After parameter control, the difference in $N_1/N_{star}$ between young and old stars shows a significant decrease from 4.6 $\sigma$ to 0.5 $\sigma$ (Figure \ref{fig:bin5_fs}).
The difference in \fkep decreases from 2.1$\sigma$ to 0.6$\sigma$.
The first bin has a higher $N_1/N_{star}$ compared to the last bin, however, it shows a slightly lower value of \fkep.
That is because the first bin has more intrinsic multi-planet systems, which can also contribute to the apparent occurrence of $N_1/N_{star}$. 
The difference in $N_{3+}/N_{star}$ between young and old stars drops moderately from 3.6$\sigma$ to 2.8$\sigma$.
Forward modeling indicates that the difference in \npbar increases slightly from 1.2$\sigma$ to 2.3$\sigma$.
Similar to the three bins case, we have only one planet system with three or more planets in the last bin, resulting in a high Poisson error ($1^{+2.3}_{-0.8}$).
As a consequence, we might underestimate the declining trend of $N_{3+}/N_{star}$.
As for $\eta$ (the product of \fkep and \npbar) the difference drops from 4.8$\sigma$ to 2.2$\sigma$, which is mainly due to the decrease of the difference in \fkep.

The Pearson correlation coefficients and $p$-values for the correlations between age and the intrinsic planet occurrence (\fkep, \npbar, and $\eta$) are also printed in the corner of each panel in Figure \ref{fig:bin5_fne}.
Regarding the anti-correlation between age and \fkep, it becomes weaker after parameter control with $p$-value rising from 0.0503 to 0.729.
The anti-correlation between age and \npbar becomes statistically significant, with $p$-value dropping from 0.0839 to 0.0052.
The anti-correlations between age and $\eta$ maintain, with $p$-value changing from 0.0021 to 0.0096.

Similar to the three bins case, $\alpha$ is not well constrained before and after parameter control.

\section{Discussions}{\label{sec:discussion}}
In this paper, we first investigate the apparent planet occurrence in terms of $N_1/N_{star}$, $N_2/N_{star}$, and $N_{3+}/N_{star}$, then from which we derive the intrinsic planet occurrence in terms of \fkep, \npbar, and $\eta$ as a function of stellar age using a forward modeling method.
We have applied a parameter control method in our analyses to remove the effects caused by other stellar properties.
We find that after parameter control, younger stars generally have higher apparent occurrence than older stars.
Specifically, the intrinsic planet occurrence in terms of the number of planets per star ($\eta$) decreases with stellar age with a confidence level of about $2\sim3\sigma$.
Such a declining trend is mainly driven by the decrease in the average multiplicity (\npbar, by about 2$\sigma$), and partially by the change in the fraction of stars with planet systems (\fkep, by less than 1$\sigma$).
In what follows, we will compare our results to those from literature and discuss the implications of these findings for our understanding of planet formation and evolution.

\subsection{Giant and Ultra Short Period Planets}\label{sec:usp}
\begin{figure}[htb!]
\centering
\includegraphics[width=.45\textwidth]{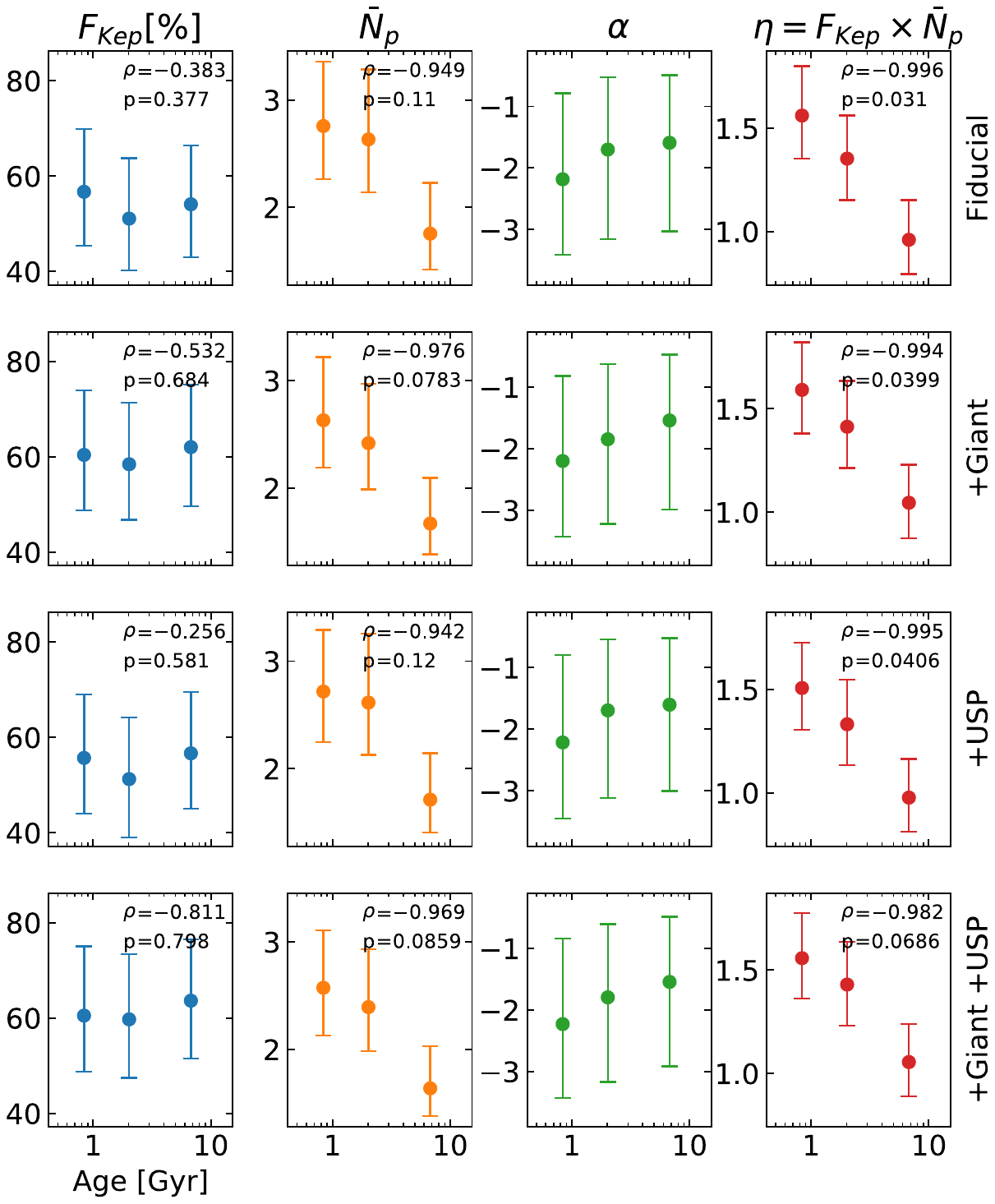}
\caption{Posterior distributions of \fkep, \npbar, $\alpha$, and $\eta$ for the three bins case are presented. From the top to the bottom, each row shows the forward modeling results after parameter control for our Fiducial planet sample, the sample including the giants, the sample including the USPs, and the sample including both of the giants and the USPs, respectively.
The dots and errorbars show the 50\% and $\pm1\sigma$ range of the posterior distributions.
In the upper right corner of the panels in the first, second, and fourth columns, we also print the Pearson correlation coefficient ($\rho$) along with the corresponding $p$-value.
\label{fig:ref_bin3}}
\end{figure}

We exclude the giant and Ultra Short Period (USP) planets from our planet sample in Section \ref{sec:data_pla}.
We dub this planet sample as the `Fiducial' sample.
To investigate the influence of giants and USPs on planet occurrence, in this section, we re-run our simulation including the giants, the USPs, and both of them.
In Figure \ref{fig:ref_bin3}, from the top to the bottom, we show the results for the Fiducial sample, the sample including giant planets, the sample including USPs, and the sample including both of giant planets and USPs for the three bins case.
As we can see, after applying the parameter control method, all results show similar trends. 
For \fkep, the differences between the youngest and oldest groups are less than 1$\sigma$. 
The youngest groups generally have \npbar $\sim2\sigma$ higher than the oldest groups, and $2\sim3\sigma$ higher for $\eta$. 
Including giant planets and USPs adds more planets into our sample, leading to a slightly higher value of \fkep. 
At the same time, since giant planets are more likely to be detected in single planet systems, including giant planets causes a very small decrease in \npbar.

\begin{figure}[htb!]
\centering
\includegraphics[width=.45\textwidth]{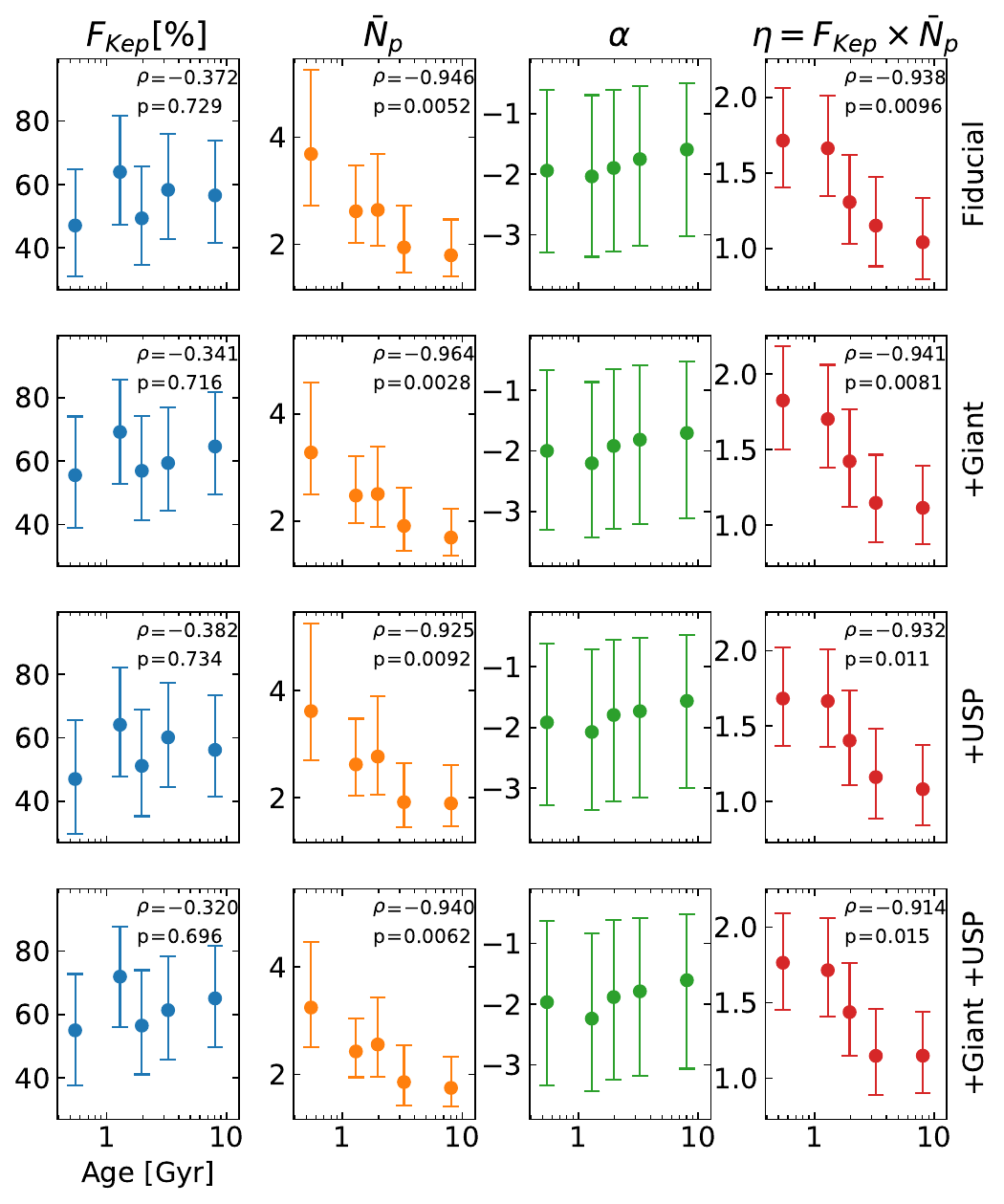}
\caption{Similar to Figure \ref{fig:ref_bin3}, posterior distributions of \fkep, \npbar, $\alpha$, and $\eta$ for five bins case are shown. 
From the top to the bottom rows, each row shows the forward modeling results corresponding to the Fiducial sample, the sample including the giants, the sample including the USPs, and the sample including both of the giants and the USPs, respectively. 
The dots and errorbars show the 50\% and $\pm1\sigma$ range of the posterior distributions.
In the upper right corner of the panels in the first, second, and fourth columns, we also print the Pearson correlation coefficient ($\rho$) and the corresponding $p$-value.
\label{fig:ref_bin5}}
\end{figure}

Similar to the three bins case, the results for the five bins case are basically unchanged after including giants and USPs.
As we can see in Figure \ref{fig:ref_bin5}, the anti-correlations between age and \fkep are statistically insignificant, with $p$-values higher than 0.5. 
For \npbar and $\eta$, the $p$-values are less than 0.05, showing significant anti-correlations between them and age.

\subsection{Comparison with Previous Studies}
\citet{McTier.2019MNRAS.489.2505M} have studied the dependence of planet occurrence rate on Galactocentric velocity.
After correcting for selection biases, they found that Kepler planet hosts have a similar velocity distribution to the non-host Kepler stars.
Based on such a similarity, they inferred that the planet occurrence rate is independent on Galactocentric velocity. 
Their inference is against our results, which show that planet occurrences in terms of \npbar and $\eta$ are anti-correlated with kinematic age and thus with Galactocentric velocity based on the Age Velocity dispersion Relation (AVR). 
In fact, we argue that their inference may not necessarily be valid for the following reasons.

First, since the occurrence rates of Kepler planets are generally found to be high \citep[$\sim$50\%,][]{Mulders.2018AJ....156...24M,Yang.2020AJ....159..164Y,He.2021AJ....161...16H}, a large fraction of the \emph{apparent} `non-host' stars are actually hosts of planets that are not detected by Kepler. Therefore, it is not surprising that Kepler planet hosts have a velocity distribution similar to that of the non-host Kepler stars as found by \citet{McTier.2019MNRAS.489.2505M}.

Second, multiple transiting systems count multiple times when counting planets, but only one time when counting host star. 
Therefore, multiples, which play a critical role in deriving the planet occurrence (e.g., \npbar) have little effect on determining the velocity distribution of host stars. 
In fact, as shown in Figure 8 of PAST II \citep{Chen.2021AJ....162..100C}, Kepler planet hosts are dominated by single transiting systems, which have a velocity distribution similar to that of non-host stars.

Our results in PAST-II have shown that multiple transiting systems have significantly different Galactocentric velocity distributions compared to the single transiting systems. 
Such differences lead to the occurrence trends with Galactocentric velocity and thus with kinematic age seen in this work (Figure \ref{fig:bin3_fne} and Figure \ref{fig:bin5_fne}, while still maintaining the similarity in Galactocentric velocity distribution between the Kepler planet hosts and the non-host Kepler stars seen in \citep{McTier.2019MNRAS.489.2505M}.
In one word, the similarity in Galactocentric velocity between Kepler planet hosts and non-host Kepler stars does not necessarily infer that the intrinsic occurrence of Kepler planets is independent of Galactocentric velocity.

In a series of papers, \citet{Bashi.2019AJ....158...61B,Bashi.2022MNRAS.510.3449B} have studied the planet occurrence rate in the Galactic context.
\citet{Bashi.2019AJ....158...61B} found that for stars with metallicity higher than $-0.25$, the planet occurrence rate generally decreases with the Galactocentric velocity of host stars.
\citet{Bashi.2022MNRAS.510.3449B} found that the planet occurrence, in terms of the fraction of stars with planets and the number of planets per star, are higher in the Galactic thin disk stars than in the thick disk stars.
In addition, \citet{Bashi.2022MNRAS.510.3449B} also showed an apparent anti-correlation between planet occurrence and the stellar isochrone age.
Generally speaking, their results are consistent with ours.
In this paper, we find that planet occurrence rate decreases with \tdd and age.
Nevertheless, we emphasize some differences in our work as compared to theirs.
First, we use the kinematic age, which has relatively smaller internal uncertainty of $\sim$10--20\% \citep{Chen.2021ApJ...909..115C} compared to isochrone age with a typical uncertainty of up to 56\% \citep{Berger_2018_ApJ_866_99B}.
Second, we use the parameter control method to isolate the effect of age from other stellar properties.
After removing these effects, we find that the anti-correlations are weaker between planet occurrence and age, especially for \fkep, though they remain significant for $\eta$ (Figure \ref{fig:bin3_fne} and \ref{fig:bin5_fne}).

\subsection{Implications to Planet Formation and Evolution}
In this study, we have revealed observational evidence that the occurrence and architecture (in terms of \fkep and \npbar) of Kepler planetary systems evolve over time.  
To gain deeper insights into planet formation and evolution, one would compare our observational results to theoretical models. Unfortunately, we did not find any models that predict \fkep or \npbar as a function of time, which would allow for a quantitative comparison with our findings. Nevertheless, there are still theoretical and numerical works in the literature that allow us to make qualitative comparisons.

Systems with more than two bodies are generally chaotic, and essentially unstable.
Planetary systems are usually formed with more than one planet, and their architecture will be further shaped during the long-term dynamical evolution afterwards, e.g., triggered by orbital instability.
The timescale of the orbital instability depends on many factors, such as mass, number, eccentricity, and orbit spacing of the planets within the system.
For a planetary system born with a large number of planets and tight orbital spacing, the orbital instability occurs quickly, which causes planet ejections and collisions, leading to a decrease in the planet number and an increase in orbital spacing \citep{Zhou.2007ApJ...666..423Z}.
This in turn increases the timescale of subsequently instability, which means the system needs to evolve on a longer time scale to trigger next instability.
As the system evolves, the instability timescale can grow to as long as billions of years.
Our Solar System may have undergone such an evolutionary process \citep{Tsiganis.2005Natur.435..459T,Liu.2022Natur.604..643L}.
In some models of our Solar System \citep[e.g.,][]{Nesvorny.2012AJ....144..117N}, it is thought there were initially five or even more giant planets formed in a tightly packed orbital configuration.
The current architecture of the Solar System was mainly shaped by an orbital instability events that ejected at least one giant and scatted other planets into a more loosely packed configuration.
For exoplanet systems, \citet{Pu.2015ApJ...807...44P} found that the orbital spacing of Kepler multi-planet systems are clustered around the threshold of orbital instability.
Based on this observation they hypothesized that most of the Kepler systems were formed with tighter spacing configuration, and most of them have undergone orbit instability, leading to fewer planets left on larger orbit spacing.
Using $N$-body simulations, \citet{Izidoro.2017MNRAS.470.1750I} proposed an evolutionary scenario for the bulk of Kepler-planet (super-Earths or mini-Neptunes) systems.
In this scenario, planets were formed in a compact resonant chains through migration in proto-planetary gas disk in the early stage.
As the gas dissipated, the chains became dynamically unstable, which led to planets merging, ejection, and being scattered to form a spread out configuration.

In this work, using a forward modeling method and after applying parameter control, we find that the planet occurrence rate in terms of planet number per star ($\eta$) decreases by about 2--3$\sigma$ as a function of time.
For three bins case, as shown in Figure \ref{fig:bin3_fne}, $\eta$ drops from 1.57 to 0.96, and for the five bins case in Figure \ref{fig:bin5_fne}, it decreases from 1.71 to 1.04.

The first major contribution to the $\eta$ decreasing trend comes from the planet number in planetary system, i.e., \npbar, since $\eta$ is the product of \npbar and \fkep.
\npbar shows a moderate decline about $\sim2\sigma$ in our fitting results.
In the bottom row of Figure \ref{fig:bin3_fne}, \npbar drops from 2.74 for stars less than 1 Gyr to 1.75 for stars about 8 Gyr, and in Figure \ref{fig:bin5_fne}, \npbar drops from 3.69 for the first age group to 1.80 for the last group.
This is qualitatively consistent with the above theories, that the dynamical evolution of planetary systems causes the merging and ejecting of planets.
Furthermore, from our results we infer that the evolution of \npbar can continue to several gigayears, which implies that planetary systems keep evolving through the whole stellar lifetime. 

The second potential contribution to the $\eta$ decreasing trend comes from the fraction of star that have planet, i.e., \fkep.
In Figure \ref{fig:bin3_fne}, \fkep changes from 56.9\% to 54.3\%, and in Figure \ref{fig:bin5_fne}, it changes from 47.1\% to 56.6\%, both are less than 1$\sigma$. 
Due to the limited star and planet sample, we cannot conclude that the change in \fkep is statistically significant.
Future studies with larger samples of planetary systems may help us to further constrain \fkep as a function of time and unveil the planet formation rate in the history of the Milky Way, combining the information on the star formation rate as a function of age \citep{Binney.2000MNRAS.318..658B}.

\begin{figure}[htb!]
\centering
\includegraphics[width=.45\textwidth]{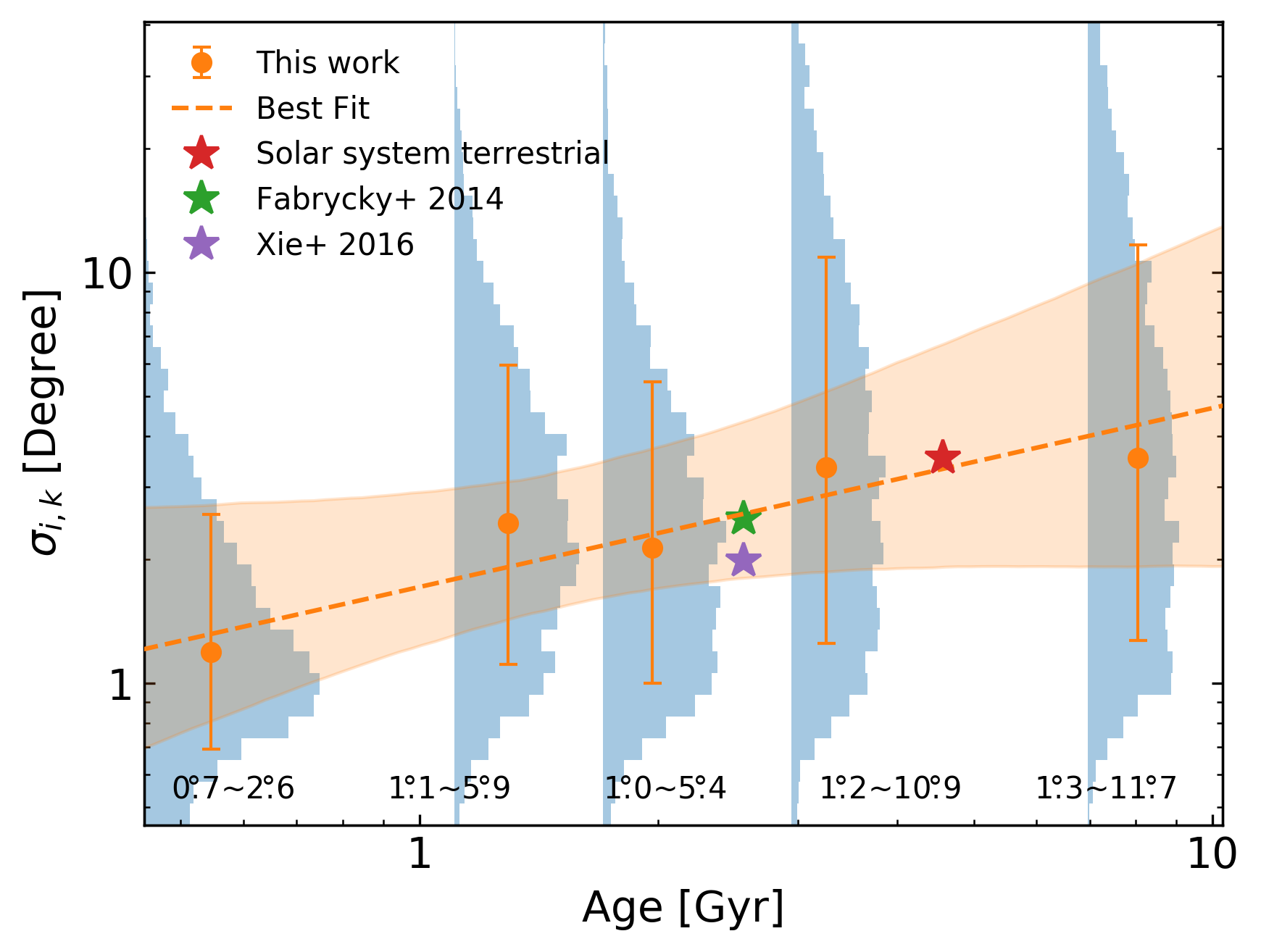}
\caption{Inclination dispersion ($\sigma_{i,k}$) as a function of age. 
We show the distribution of $\sigma_{i,k}$ for five age bins after parameter control with blue bars. 
The median value and the $\pm1\sigma$ percentile range of $\sigma_{i,k}$ are indicated by orange dots and errorbars, with the corresponding $\pm1\sigma$ percentile range is displayed in the lower part of the plot. 
The orange dashed line represents the best fit, and the orange shaded region denotes the corresponding uncertainties of the fit by resampling $\sigma_{i,k}$ 10,000 times.
The $\sigma_{i,k}$ value of the Solar System terrestrial planets is shown with the red star marker, and the $\sigma_{i,k}$ values of Kepler multiple transiting systems measured by \citet{Fabrycky.2014ApJ...790..146F} and by \citet{Xie_2016_PNAS_11311431X} are shown with the green and purple star markers, respectively.
The age of the Solar System is adopted as 4.57 Gyr \citep{Bouvier.2010NatGe...3..637B}, and the age of Kepler multiples is adopted as the kinematic age of stars that host two or more planets in our star sample.
\label{fig:bin5_sik}}
\end{figure}

Not only does the number of planets in a system evolve with time, but the orbital properties also undergo changes.
Since \npbar is related to the orbital inclination (Equation \ref{eq:inclination}), we can investigate the mutual orbital inclination as a function of time.
We calculate the posterior distributions of $\sigma_{i,k}$ for the five bins after parameter control, and show the distributions as well as the median value and 1$\sigma$ range of $\sigma_{i,k}$ in Figure \ref{fig:bin5_sik}.
As we can see, $\sigma_{i,k}$ gradually evolves with time.
From less than 1 Gyr to about 8 Gyr, the median value of $\sigma_{i,k}$ grows from about $1\fdg2$ to $3\fdg5$, and the $1\sigma$ range expands from $0\fdg7$--$2\fdg6$ to $1\fdg3$--$11\fdg7$.
To further quantify the age-$\sigma_{i,k}$ trend, we fit $\sigma_{i,k}$ with age.
Although bearing large uncertainty (as seen from the orange shaded region), the best fit is
\begin{equation}
\log{\sigma_{i,k}}=0.2+0.4\times\log{\frac{\text{Age}}{\text{Gyr}}}.
\end{equation}
For comparison, we also plot the data points for our Solar System and Kepler multiple transiting systems in Figure \ref{fig:bin5_sik}.
They are all generally fit such an age-$\sigma_{i,k}$ trend.
This result indicates that as planetary systems get older, they become dynamically hotter, which is consistent with the theoretical expectation \citep{Zhou.2007ApJ...666..423Z}.
The Kepler multiples show smaller mutual inclinations compared to our Solar System, which can be explained by their younger ages according to the age-inclination trend shown in Figure \ref{fig:bin5_sik}.
In other words, Figure \ref{fig:bin5_sik} may hint that planets in our Solar System were in a flatter architecture in the early time, then gradually evolve to the current state.

\section{Summary and Conclusion}{\label{sec:conclusion}}
In this work, which is the fourth paper of the PAST series, we update the LAMOST-Gaia-Kepler catalog utilizing the recently released LAMOST DR8 and Gaia DR3.
Based on this catalog, we study the occurrence rate and architecture of Kepler-like planets as a function of stellar kinematic age.
We find the following results.
\begin{enumerate}[{(}1{)}]
    \item Younger stars generally show higher apparent planet occurrence rates (more than 3$\sigma$) for one, two, and three or more planets ($N_1/N_{star}$, $N_2/N_{star}$, and $N_{3+}/N_{star}$) than older stars (top rows of Figure \ref{fig:bin3_fs} and Figure \ref{fig:bin5_fs}).
    
    \item Applying a parameter control method can effectively reduce the effects caused by other stellar properties, such as effective temperature, mass, metallicity, radius, and $\sigma_\text{CDPP}$.
    After parameter control, the differences in $N_1/N_{star}$ and $N_2/N_{star}$ between younger and older stars decrease to less than 2$\sigma$, while the difference in $N_{3+}/N_{star}$ maintains a confidence level of about 3$\sigma$ (bottom rows of Figure \ref{fig:bin3_fs} and Figure \ref{fig:bin5_fs}). 
    
    \item Adopting a forward modeling method can help us to investigate the intrinsic planet occurrence in terms of the fraction of stars with planetary systems (\fkep), the average planet multiplicity (\npbar), and the number of planets per star ($\eta$).
    For stars without parameter control, we find that the younger stars have higher \fkep and \npbar by about 2$\sigma$ than older stars.
    The difference in $\eta$ between younger and older stars is more obvious, at about 5$\sigma$ level (top rows of Figure \ref{fig:bin3_fne} and Figure \ref{fig:bin5_fne}).
    
    \item After parameter control, the differences in \fkep drops to less than 1$\sigma$, hinting the planetary system occurrence remains at a similar rate throughout the history of the Milky Way.
    The difference in \npbar is about 2$\sigma$ between the younger and older stars.
    This result is consistent with theories that planet systems keep evolving as a result of the merging and ejecting of the planets.
    Younger stars have a higher $\eta$ (the product of \fkep and \npbar) by about $2\sim3\sigma$ than older stars, which is the combining effects caused by the evolution of \fkep and \npbar (bottom rows of Figure \ref{fig:bin3_fne} and Figure \ref{fig:bin5_fne}). 
    
    \item The orbital properties of planet systems also evolve with time. 
    We find that in stars aging from less than 1 Gyr to about 8 Gyr, the mutual orbital inclination ($\sigma_{i,k}$) between their planets increases from $1\fdg2$ to $3\fdg5$, and the $\pm1\sigma$ range of $\sigma_{i,k}$ expands from $0\fdg7$--$2\fdg6$ to $1\fdg3$--$11\fdg7$ (Figure \ref{fig:bin5_sik}), hinting that planet systems become dynamically hotter as a function of time.
    Both our Solar System and Kepler multiple transiting systems fit such a trend.
    
\end{enumerate}

Our work qualitatively agrees with theoretical expectations that planet occurrence decreases, and planetary systems become dynamically hotter with age. 
Future dedicated theoretical and numerical modeling on the occurrence and architecture of Kepler planets as a function of age are needed to allow us to make quantitative comparisons to our results in this work and to place key constraints on planet formation and evolution.

The current and upcoming missions also aid in exploring exoplanets in the dimension of time.
The TESS mission has found thousands of candidates \citep{Guerrero.2021ApJS..254...39G}, covering a wide range of ages \citep[e.g.,][]{Newton.2019ApJ...880L..17N,Gan.2020AJ....159..160G,Weiss.2021AJ....161...56W}.
In this paper, our studies only rely on a portion of planets from the Kepler sample.
Both the sample size and the number of bins are still limited, leading to relatively large uncertainties in our fitting results.
In the near future, missions such as Gaia, TESS \citep{Ricker.2015JATIS...1a4003R}, and PLATO \citep{Rauer.2014ExA....38..249R} will detect many more exoplanets, leading to the expansion of planet sample by one order of magnitude or even more.
With the help of more data, future studies will further refine our measurements and test our results.

\acknowledgments
This work is supported by the National Key R\&D Program of China (No. 2019YFA0405100) and the National Natural Science Foundation of China (NSFC; grant Nos. 12150009, 11933001, 12273011). J.-W. X. also acknowledges the support from the National Youth Talent Support Program and the Distinguish Youth Foundation of Jiangsu Scientific Committee (BK20190005). D.-C.C. also acknowledges the Cultivation project for LAMOST Scientific Payoff and Research Achievement of CAMS-CAS.

This work has included data from Guoshoujing Telescope (the Large Sky Area Multi-Object Fiber Spectroscopic Telescope, LAMOST), which is a National Major Scientific Project built by the Chinese Academy of Sciences.
Funding for LAMOST (www.lamost.org) has been provided by the Chinese NDRC. LAMOST is operated and managed by the National Astronomical Observatories, CAS.
This work presents results from the European Space Agency (ESA) space mission Gaia. Gaia data are being processed by the Gaia Data Processing and Analysis Consortium (DPAC). Funding for the DPAC is provided by national institutions, in particular the institutions participating in the Gaia MultiLateral Agreement (MLA).

\software{emcee \citep{Foreman-Mackey.2013PASP..125..306F}, scikit-leart \citep{Pedregosa.2012arXiv1201.0490P}, KeplerPORTs \citep{Burke.2017ksci.rept...19B}, numpy \citep{Harris.2020Natur.585..357H}, matplotlib \citep{Hunter.2007CSE.....9...90H}, astropy \citep{Astropy.2013A&A...558A..33A,Astropy.2018AJ....156..123A,Astropy.2022ApJ...935..167A}, evolstate \citep{Huber.2019ascl.soft05003H}, RGCCA \citep{girka.rgcca}}

\newpage
\appendix

\section{Updating kinematic Characteristics of Galactic components and Age-velocity dispersion relation with Gaia DR3 astrometry data}{\label{sec:app_edr3}}
In the first paper of the PAST series \citep[PAST \RNum{1}, ][]{Chen.2021ApJ...909..115C}, we revisited the kinematic method to classify the Galactic components and the Age-Velocity dispersion Relation (AVR) with Gaia DR2 \citep{2018A&A...616A...1G,2018A&A...616A..11G} and the LAMOST main-sequence turn-off and subgiant  (MSTO-SG)  star sample \citep{2017ApJS..232....2X}.
On June 13th, 2022, Gaia Data Release 3 \citep[DR3;][]{Gaia.2022arXiv220800211G} was released, providing five astrometric parameters (positions, parallaxes, and proper motions) for 1.468 billion sources. 
Comparing to Gaia DR2, the standard uncertainties have been reduced for the positions, parallaxes, and proper motions, which make the astrometric results considerably more robust and reduce the systematic errors. 
Therefore, here we revisit the kinematic methods and AVR with Gaia DR3 by adopting the same procedures shown in the Section 2 and 3 of PAST \RNum{1}.

\subsection{Updated Calibration Sample}
To construct the calibration sample, we first cross-match the above LAMOST MSTO-SG catalog with the Gaia DR3 catalog using the X-match service provided by the Centre de Donnees astronomiques de Strasbourg (CDS, http://cdsxmatch.u-strasbg.fr).
Second we carry out a angular distance cut of 1.25 arcseconds and a Gaia G-band magnitude difference cut of 2.5 mag.
For stars with multiple matches, we keep those with the smallest angular separation.

Then we calculate the stellar kinematic properties (i.e., Galactocentric cylindrical coordinates $(R, \theta, Z)$, and Galactic rectangular velocities $(U_{\rm LSR}, V_{\rm LSR}, W_{\rm LSR})$) relative to the local standard of rest (LSR), with the procedure detailed in Section 2.1 of PAST \RNum{1}.
We adopt the location of the Sun of $R_\odot$= 8.18 kpc \citep{2019A&A...625L..10G,2021A&A...649A...1G} and $Z_\odot$ = 25 pc \citep{2001ApJ...553..184C}. 
The solar peculiar motions are taken as [$U_\odot$, $V_\odot$, $W_\odot$] = [9.58, 10.52, 7.01] $\rm km \ s^{-1}$  \citep{2015ApJ...809..145T}.

After that, we apply the following filters to further clean the calibration sample.  

  \  (1) Binary filter. We remove binary systems because their kinematics contain additional motions \citep{1998MNRAS.298..387D}. This is done by choosing stars flagged as `Normal star' (i.e., single stars with spectral types of AFGKM) in the LAMOST MSTO-SG catalog \citep{2017ApJS..232....2X}. 
  We also remove potential binaries by eliminating stars with Gaia DR3 re-normalized unit-weight error (RUWE) $>1.4$ \citep{LL:LL-124}.
  
  \ (2) Parallax precision filter. Following \cite{1998MNRAS.298..387D}, we remove stars with relative parallax errors larger than 10 percent as reported in the Gaia DR3.
  
  \  (3) Age precision filter. We remove stars with ages older than 14 Gyr, errors of age exceeding 25\%, or blue straggler stars ($|Z|> 1.5$ kpc and ages younger than 2 Gyr) in the LAMOST MSTO-SG catalog.
  
  \ (4) Distance filter \citep[similar to][]{1997ESASP.402..473B}.  The majority of the remaining stars are brighter than G mag=16, where the median parallax error is 0.0494 mas. Recalling the above 10 percent parallax precision requirement, this translates to a distance limit  $\sim 1/(0.0494/0.1) \sim 2.0$ kpc. 
  We therefore remove stars with distances exceeding than this limit.
  
After applying the above filters, we are left with 134,244 stars, which are mainly (129,089/134,244, 96.2\%) located at  $7.5 < R < 10.0$ kpc, $|\theta| < 15$ deg, and $|Z|<1.5$ kpc. 

\subsection{Revisiting the Kinematic Method To Classify the Galactic Components}
With stellar kinematic and age following the same criteria in the Section 2.3.3 of PAST \RNum{1}, we then classify the calibration sample into different Galactic components, i.e., thin disk ($D$), thick disk ($TD$), halo ($H$), and Hercules stream ($Herc$).
In order to calculate the characteristic kinematic parameters for each Galactic component as a function of $(R, \ Z)$, we bin the calibration sample as the same interval in PAST \RNum{1}.
For $|Z|$, we set 8 bins with boundaries at $|Z|=$ 0, 0.1, 0.2, 0.3, 0.4, 0.55, 0.75, 1.0, and 1.5 kpc.
For $R$, we set 5 bins with boundaries at $R =$ 7.5, 8.0, 8.5, 9.0, 9.5, and 10 kpc. 
In total, there are $5 \times 8=40$ grids in the $R$--$Z$ space, and all bins have enough ($>400$) stars.

\begin{figure*}[!htb]
\centering
\includegraphics[width=\textwidth]{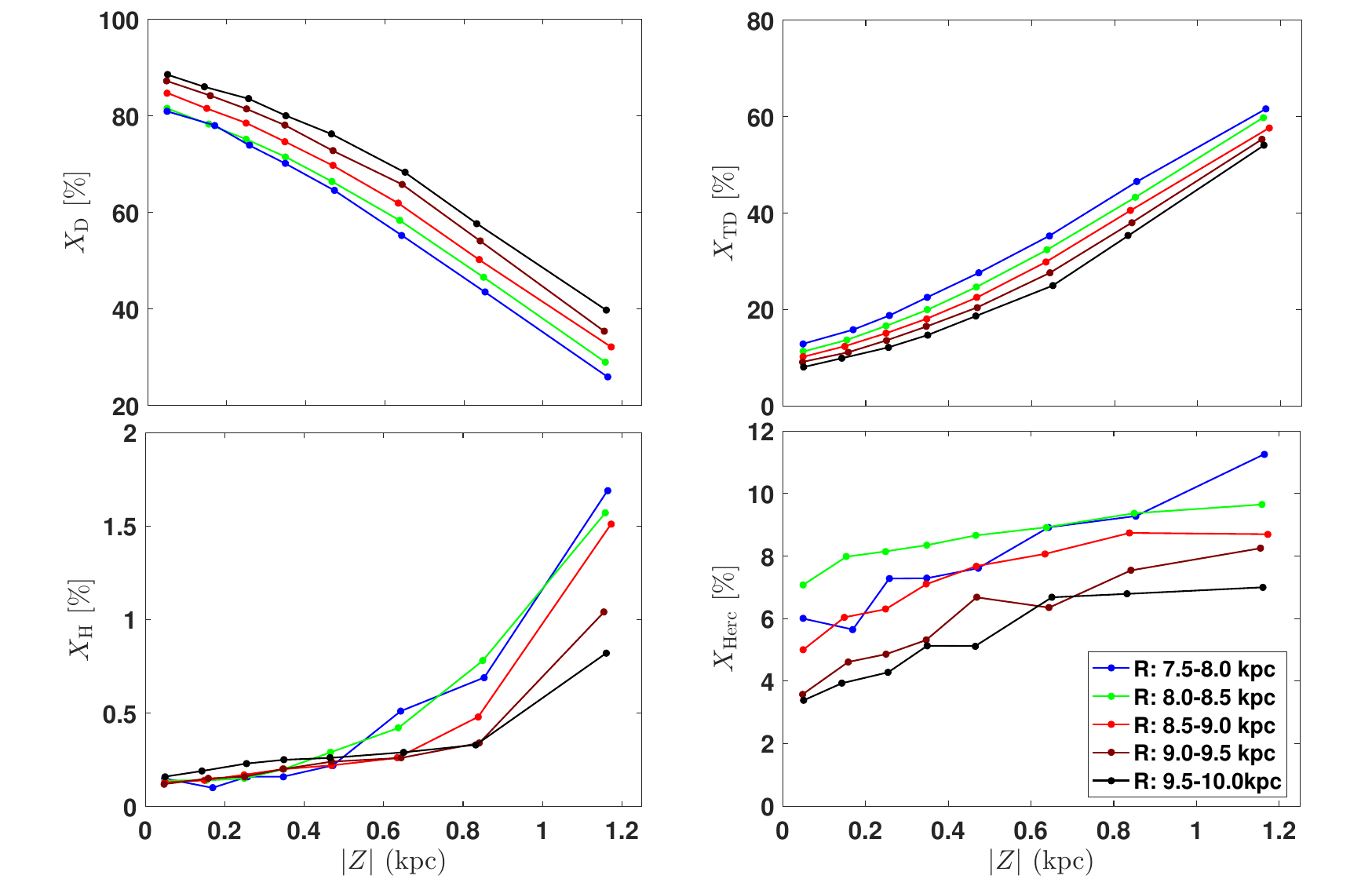}
\caption{The normalisation fraction $X$ of stars for each component as functions of {Galactic} radius $R$ and absolute value of height $|Z|$. The different colours denote subsamples of stars with different Galactic radii.
\label{fXDTDHHerGaiaeDR3MOST}}
\end{figure*}

\begin{figure*}[!t]
\centering
\includegraphics[width=0.9\textwidth]{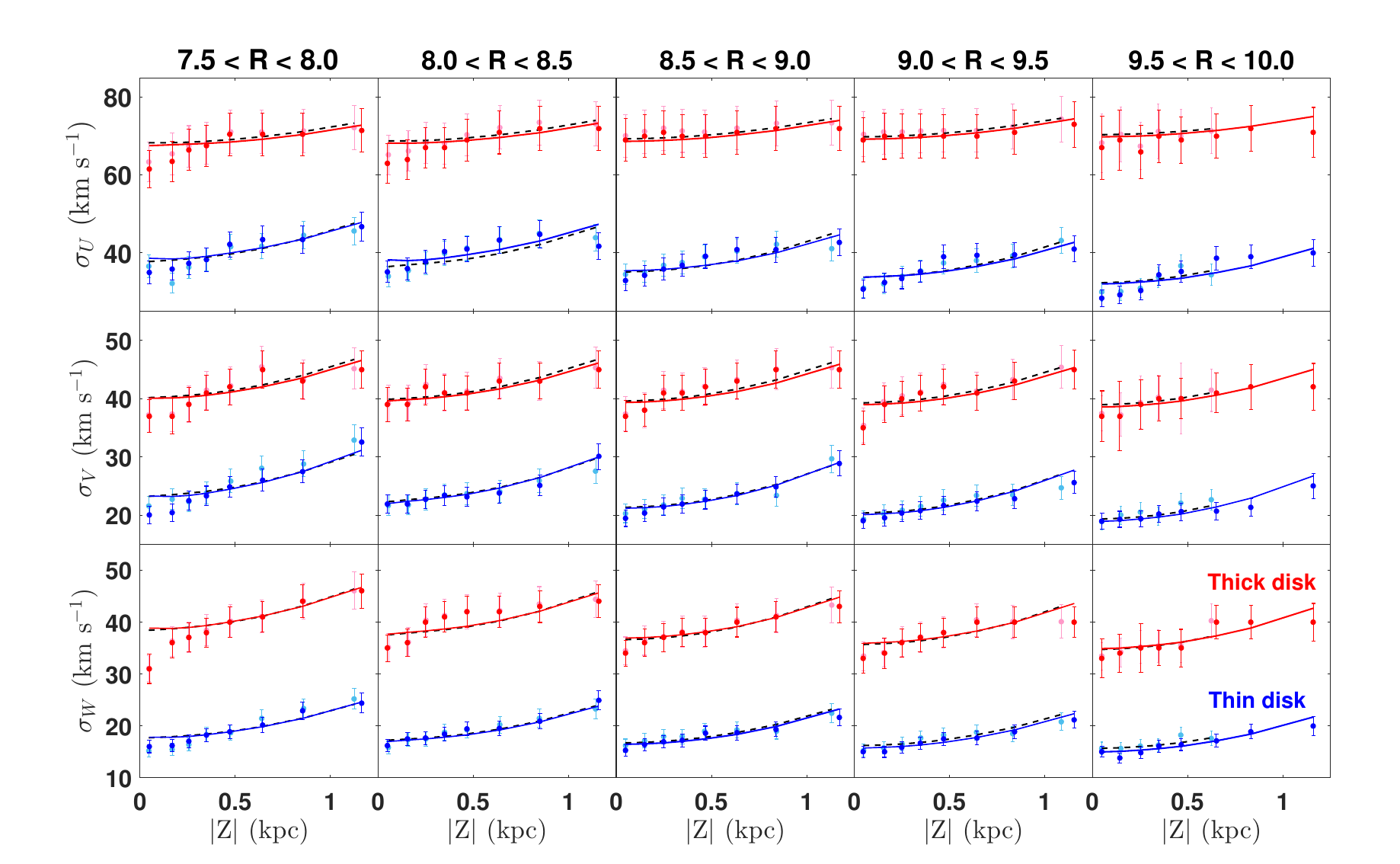}
\caption{The velocity dispersions as functions of position ($R$, $|Z|$) in the Galaxy. 
For the updated calibration data conducted from the LAMOST MSTO-SG and Gaia DR3 catalogs, the velocity dispersions are plotted as solid points, and the line segments represent $1\sigma$ errors in two colors: blue for thin disk and red for thick disk.
The solid line in each panel denotes the result of the best fit of Equation \ref{eUVWRZ} using the coefficients in Table \ref{tab:eUVWparaGaiaeDR3MOST}.
While for the calibration sample conducted from the LAMOST MSTO-SG and Gaia DR2 catalogs in the PAST \RNum{1}, the velocity dispersions are plotted as solid points, and the line segments represent $1\sigma$ errors in two colors: light blue for thin disk and light red for thick disk.
The dashed black line in each panel denotes the result of the best fit of Equation \ref{eUVWRZ} using the coefficients in Table \ref{tab:eUVWparaGaiaeDR3MOST}.
\label{figeUVWRZGaiaeDR3MOST}}
\end{figure*}

\begin{figure}[!t]
\centering
\includegraphics[width=0.8\textwidth]{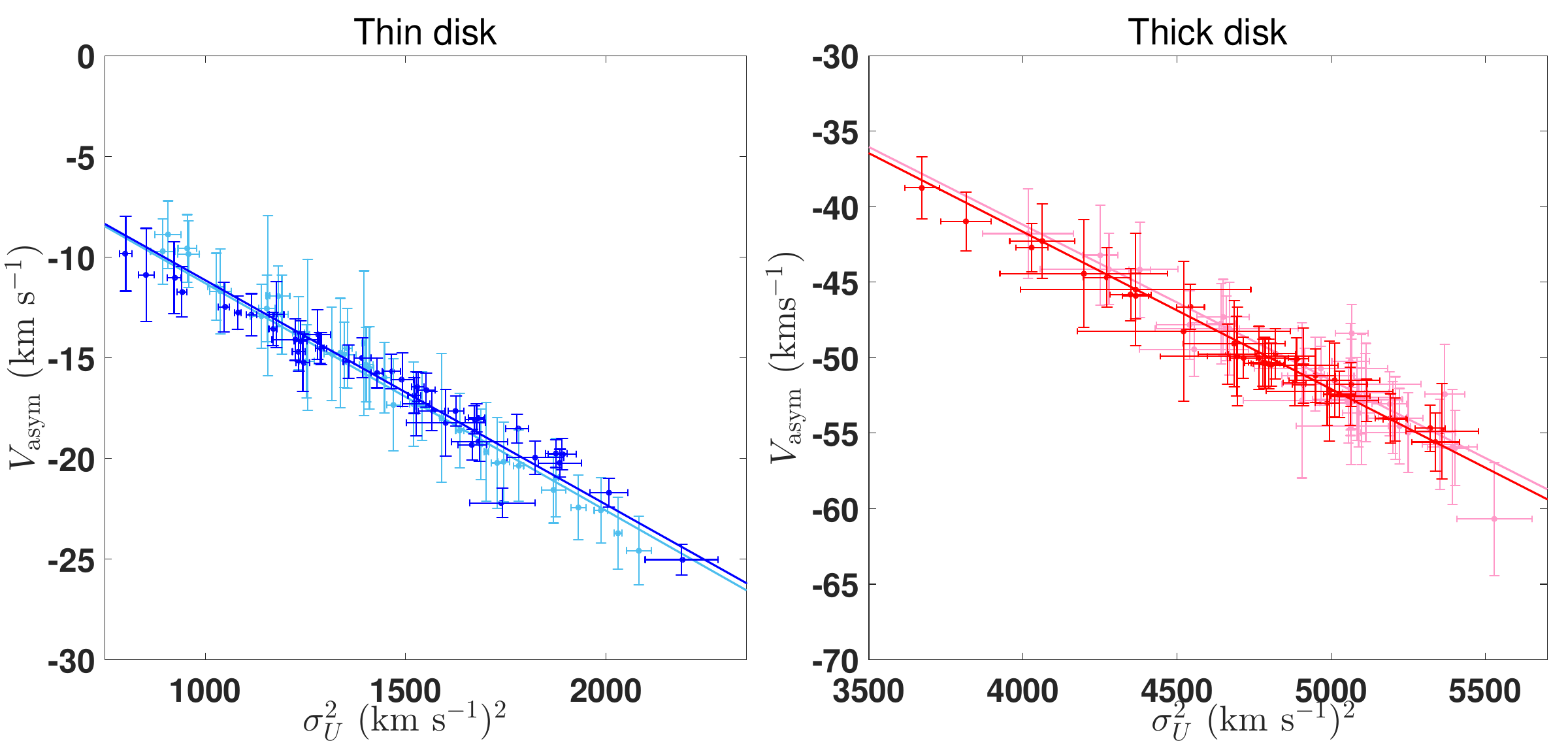}
\caption{The asymmetric velocity, $V_{\rm asym}$ as a function of $\sigma_U^2$ for the thin disk (left panel) and thick disk (right panel). 
For the updated calibration data conducted from the LAMOST MSTO-SG and Gaia DR3 catalogs, the data is plotted as blue/red points and the blue/red line segments represent $1\sigma$ errors.
The blue/red solid lines denote the results of the best fit using Equation \ref{eqVasym}. 
While for the calibration sample conducted from the LAMOST MSTO-SG and Gaia DR2 catalogs in the PAST \RNum{1}, the data and $1\sigma$ errors are plotted in light blue/red color.
The light blue/red dashed lines denote the results of the best fit using Equation \ref{eqVasym}.
\label{figVasym}}
\end{figure}

We then revise the normalized fraction $X$ (Equations 9, 10, 11, 12 in PAST \RNum{1}) and the velocity ellipsoid (i.e., $\sigma_U$, $\sigma_V$, $\sigma_W$, and $V_{\rm asym}$ (Equation 3 in PAST \RNum{1}) of each Galactic component for each grid in the $R$--$Z$ plane following the same procedure of Section 2.3.4 and 2.3.5 of PAST \RNum{1}.
The calculated values of $X$, $\sigma_U$, $\sigma_V$, $\sigma_W$, and $V_{\rm asym}$ are tabulated in Table \ref{tab:dispersionsrevised} and visualized in Figure \ref{fXDTDHHerGaiaeDR3MOST}, Figure \ref{figeUVWRZGaiaeDR3MOST}, and Figure \ref{figVasym}.

Figure \ref{fXDTDHHerGaiaeDR3MOST} shows the $X$ values of various Galactic components as functions of Galactic radius $R$ and absolute value of height, $|Z|$.
As expected, $X_{\rm D}$ ($X_{\rm TD}, \ X_{\rm H}$) generally decreases (increases) with $|Z|$ in all the $R$ bins.

With the same procedure detailed described in Section 2.3.4 of PAST \RNum{1},  we also fit the velocity dispersions, $\sigma_U$, $\sigma_V$, and $\sigma_W$ in the following formula according to \cite{2013MNRAS.436..101W}:
\begin{equation}
\sigma = b_1 + b_2\times \frac{R}{\rm kpc} + b_3 \times (\frac{Z}{\rm kpc})^2 {\rm km \ s^{-1}}.
\label{eUVWRZ}
\end{equation}
We then use the following formula to calculate $V_{\rm asym}$ according to \citet{2003A&A...409..523R,2008gady.book.....B}
\begin{equation}
V_{\rm asym} = \sigma_U^2/C_0.
\label{eqVasym}
\end{equation}
The values of fitting parameters and their 1$\sigma$ uncertainties are summarized in Table \ref{tab:eUVWparaGaiaeDR3MOST}.

Comparing to the results obtained from Gaia DR2 and LAMOST MSTO-SG sample in PAST \RNum{1}, we find that for the normalized fraction $X$,  the typical (median) relative differences are only 0.6\%, -3.5\%, 4.6\%, and 5.2\% for the thin disk, thick disk, halo, and Hercules stream, respectively. 
For the velocity ellipsoid (i.e., $\sigma_U$, $\sigma_V$, $\sigma_W$, and $V_{\rm asym}$) obtained with the calibration sample using astrometry data from Gaia DR2 and DR3, as can be seen in Figure \ref{figeUVWRZGaiaeDR3MOST} and \ref{figVasym}, the median values and $1\sigma$ errorbars are very similar, and the best fits are nearly the same with each other.
It can also be seen from Table \ref{tab:AVRkbGaiaeDR3MOST}, the fitting parameters (i.e., $b_1$, $b_2$, $b_3$, and $C_0$) are well consistent with those of PAST \RNum{1} within their $1\sigma$ errorbars.
Therefore, we conclude that the $X$ factors and velocity ellipsoid obtained with the updated calibration sample are broadly unchanged from those of PAST \RNum{1}.

\subsection{Revisiting the Age-Velocity dispersion relation (AVR)}
According to PAST \RNum{1}, we divide the calibration sample into 30 bins with approximately equal sizes ($\sim$ 4,475 stars in each bin) according to their ages. 
Then we fit the AVRs following \citet{2009A&A...501..941H,2016MNRAS.462.1697A} by using a simple power law formula, i.e.,
\begin{equation}
\sigma =  k \times \left(\frac{t}{\rm Gyr} \right)^{\beta} \, \rm km \ s^{-1},
\label{eqAVR}
\end{equation}
where $t$ represents stellar age, $\sigma$ is the velocity dispersion, and $k$ and $\beta$ are two fitting parameters.
The best fits and uncertainties ($1\sigma$ interval) of the fitting parameters $(k, \ \beta)$ are calculated with the same procedure described in Section 3.1 of PAST \RNum{1} and summarized in Table \ref{tab:AVRkbGaiaeDR3MOST}.

\begin{figure*}[!t]
\centering
\includegraphics[width=\textwidth]{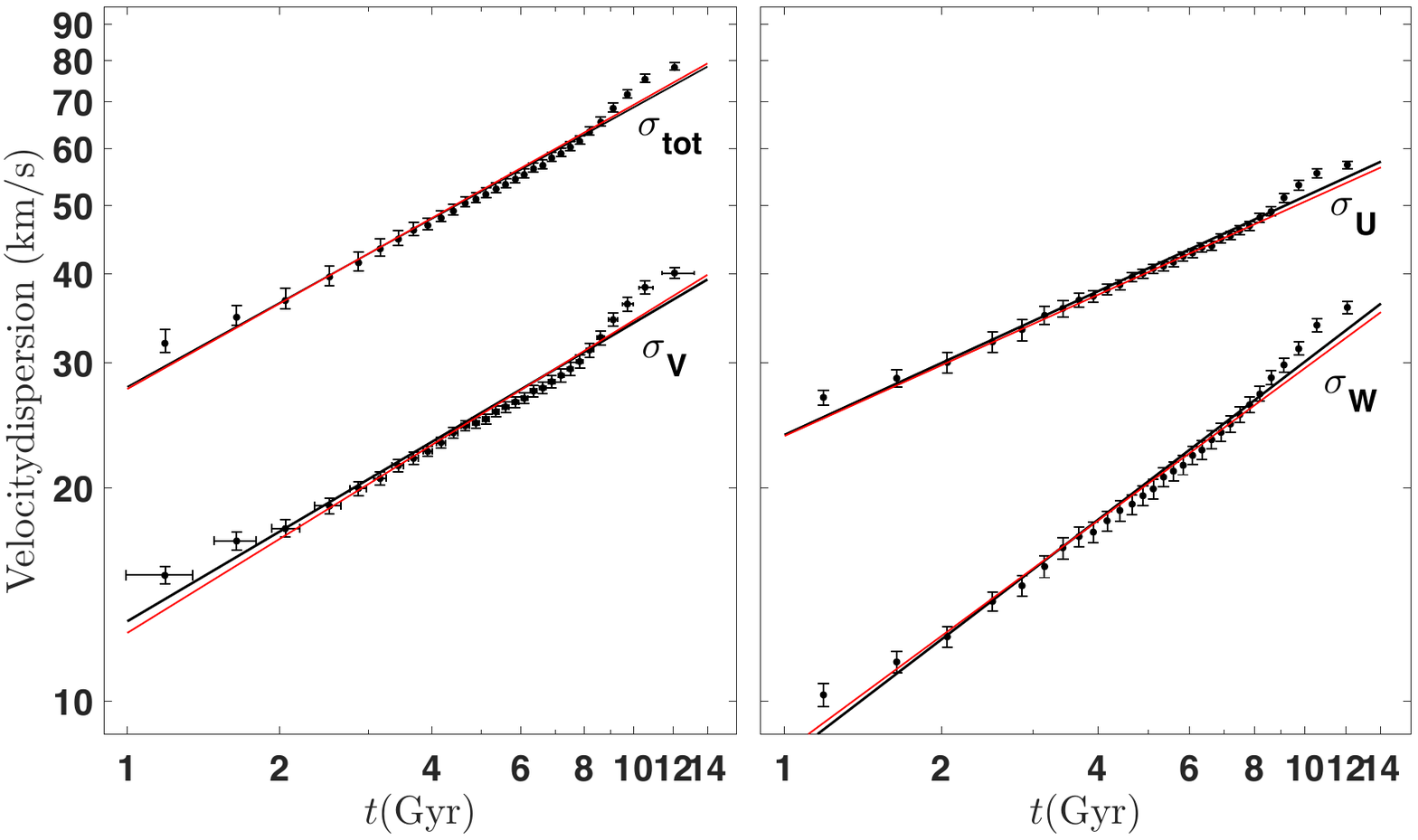}
\caption{The velocity dispersions for $U_{\rm LSR}, V_{\rm LSR}, W_{\rm LSR}$, and $V_{\rm tot}$ vs. age for the selected calibration star sample.
The solid black lines denote the respective best fit of refitting AVR (Equation \ref{eqAVR}) using the coefficients in Table \ref{tab:AVRkbGaiaeDR3MOST}.
\label{fAVRGaiaeDR3}}
\end{figure*}

Figure \ref{fAVRGaiaeDR3} shows the velocity dispersion as a function of the median age of each bin.
As can be seen, the best fits (black lines) for the relationship between age and the dispersion of velocity components ($U_{\rm LSR}, V_{\rm LSR}, W_{\rm LSR}$) and the total velocity ($V_{\rm tot}$) are all indistinguishable from those of PAST \RNum{1} (red lines).
In Table \ref{tab:AVRkbGaiaeDR3MOST}, we compare the fitting parameters $(k, \ \beta)$ of AVRs obtained from the updated calibration sample to those from PAST \RNum{1}. As can be seen, the median values are nearly identical, and the $1\sigma$ uncertainties of $k$ decrease by a factor of $\sim 10\%$ due to the improvement on the precision of stellar parallax and proper motion measurements.
Thus, the AVRs derived with calibration sample using astrometry data from Gaia DR3 and DR2 are nearly the same with each other.

\section{Deriving the Age-Planet Occurrence Relationship through Canonical Correlation Analysis}\label{sec:cca}
Canonical Correlation Analysis (CCA) is an effective method to discover the correlations between different sets of variables.
It was first introduced by \citep{HOTELLING.10.1093/biomet/28.3-4.321}.
The basic idea is to identify linear combinations of two sets of variables that the resulting combined variables exhibit the highest possible correlation.
\citet{KETTENRING.10.1093/biomet/58.3.433} summarized various methods for establishing connections among multiple sets of variables.
Here, we apply the CCA method to investigate the relationship between planet and star properties.
Specifically, we use the SABSCOR method to maximize the sum of the absolute values of the correlation among different sets.

We group the planet and star properties into three sets of variables: the planet system property (occurrence rate), the interesting stellar property (kinematic age), and the stellar properties we want to eliminate (effective temperature, mass, metallicity, radius, and $\sigma_{\text{CDPP}}$), respectively.

The planet system property we are interested in is the apparent occurrence rate. 
Following \citet{Zhu.2019ApJ...873....8Z}, we use two tracers to represent the occurrence rate, which are defined by the following equations:
\begin{align}
     \text{Tracer}(\eta) &=\frac{1}{N_{star}} \sum_{j=1}^K j N_j\label{eq:teta}\\
     \text{Tracer}({F_p}) &=\frac{1}{N_{star}} \left( N_1 + \sum_{j=1}^K N_j \right).\label{eq:tfp}
\end{align}
Tracer($\eta$) is related to the average number of planets per star, and Tracer($F_p$) is correlated with the fraction of stars processing planet systems. 
In these equations, $N_{star}$ represents the number of stars in each bin, $N_j\ (j=1,2...K)$ is the number of systems with $j$ planets, and $K$ is the maximum number of planets we observed in the planetary system. 
Given the transit method can only detect a very small fraction of planets, in our sample of 19,358 stars, we have observed only 641 planets in 467 planetary systems. 
Therefore, to calculate these two tracers, we need to group stars into bins. 
Additionally, the distribution of planet systems is not uniform, especially for systems with three or more planets, with fewer systems found around older stars. To reduce Poisson errors, it is necessary to limit the number of bins. 
Consequently, we group the stars into 40 bins with an equal number of stars (using 30 or 50 bins yields similar results).

The interesting stellar property, kinematic age, is determined using the Age-Velocity dispersion Relationship (AVR, see Appendix \ref{sec:app_edr3}).
This relationship requires the calculation of velocity dispersion and, as a result, is applicable only to a group of stars. 
Therefore, we once again need to divide our star sample into bins.

For the uninteresting stellar properties, we use the median value of each bin as the representative value.
Given the limited number of bins (40), to prevent overfitting, we select only three properties (mass, [Fe/H], and $\sigma_{\text{CDPP}}$) instead of all five. 
Previous studies \citep[e.g.,][]{Yang.2020AJ....159..164Y,He.2021AJ....161...16H} have shown that the mass and effective temperature of stars have a similar influence on planet occurrence. From late type to early type stars, the increase in mass and temperature leads to a decrease in planet occurrence. 
We choose mass to represent the influence of stellar type. 
Furthermore, an increase in radius and $\sigma_{\text{CDPP}}$ both result in a reduction in detection efficiency, which lowers the probability of planet detection. 
Here we choose $\sigma_{\text{CDPP}}$ to present the effect of detection efficiency (choosing temperature or radius leads to a similar result).

In summary, we categorize all the planet/star properties into three groups. 
The first group, $\mathbf{X}_1$=\{Tracer($\eta$), Tracer($F_p$)\}, represents planet occurrence. 
The second group, $\mathbf{X}_2$=\{Age\}, describes the interesting stellar properties, specifically, stellar kinematic age. 
The third group, $\mathbf{X}_3$=\{Mass, [Fe/H], $\sigma_{\text{CDPP}}$\}, is related to uninteresting stellar properties.

We use the R package RGCCA \citep{girka.rgcca}, to maximize the sum of the correlations between planet occurrence and stellar age, as well as between planet occurrence and uninteresting star properties.
This can be formulated as solving the following optimization problem:
\begin{equation}
    \text{Maximize}\ (|\text{Cor}(\mathbf{y}_1,\mathbf{y}_2)|+|\text{Cor}(\mathbf{y}_1,\mathbf{y}_3)|).
\end{equation}
where $\mathbf{y}_1=\mathbf{X}_1\mathbf{a}_1$, $\mathbf{y}_2=\mathbf{X}_2\mathbf{a}_2$, and $\mathbf{y}_3=\mathbf{X}_3\mathbf{a}_3$, and $\mathbf{a}_1$, $\mathbf{a}_2$, and $\mathbf{a}_3$ are the weight vectors for each variable set.
All the variables have been standardized.
The results can be seen in the following figure.

In Figure \ref{fig:cca}, each ellipse represents a group, and each box represents a star/planet property. 
We also print the weight and correlation on each line. 
As we can see, both Tracer($\eta$) and Tracer($F_p$) have positive contributions to the planet occurrence ($\mathbf{a}>0$), showing that as the number of planets per star and the fraction of stars with planets increase, the planet occurrence rises. 
The kinematic age is in anti-correlated with planet occurrence (Cor=-0.802), which is consistent with our result before parameter control (see top rows of Figure \ref{fig:bin3_fs} and Figure \ref{fig:bin5_fs}). 
As stars aging from less than 1 Gyr to about 8 Gyr, both of the fraction of stars with planetary systems and the number of planets per star decrease. 
The uninteresting star properties are correlated with planet occurrence. 
Among these properties, the stellar mass shows a negative weight, which is in agreement with literature \citep[e.g.][]{Yang.2020AJ....159..164Y,He.2021AJ....161...16H}.
An increase in mass results in a decrease in Kepler-like planet occurrence. 
Metallicity shows a positive weight, which is generally consistent with previous studies \citep[e.g.,][]{Zhu.2019ApJ...873....8Z,Wang.2015AJ....149...14W}, indicating that an increase in metallicity can stimulate the formation of planets. 
$\sigma_{\text{CDPP}}$ demonstrates a negative weight because higher $\sigma_{\text{CDPP}}$ leads to lower detection efficiency, which hinders the detection of planets.

The CCA method shows a similar result to ours before parameter control, indicating that the increase in stellar age results in a decrease in the planet occurrence rate. 
This decrease is reflected in both the fraction of stars with planetary systems and the number of planets per star.

\section{Detection Efficiency}{\label{sec:de}}
We show the average detection efficiencies and planet samples for both the three and five bins cases in Figure \ref{fig:bin3_de} and \ref{fig:bin5_de}.
The detection efficiency metrics are calculated by the Package \texttt{KeplerPORTs} \citep{Burke.2017ksci.rept...19B}, and the associated data are downloaded from NASA exoplanet archive\footnote{https:// exoplanetarchive.ipac.caltech.edu/docs/}.

As we can see in the top rows of Figure \ref{fig:bin3_de} and \ref{fig:bin5_de}, young stars have slightly higher detection efficiencies than old stars.
This is because first, young stars generally have smaller stellar radii, which lead to deeper transit depths, and second, young stars have lower noise levels ($\sigma_\text{CDPP}$) that increase the signal to noise ratio (top rows of Figure \ref{fig:bin3_cdf} and Figure \ref{fig:bin5_cdf}).
After we apply parameter control to remove the effects caused by stellar properties, the young and old stars have similar distributions of stellar radii and $\sigma_\text{CDPP}$ (bottom rows of Figure \ref{fig:bin3_cdf} and Figure \ref{fig:bin5_cdf}).
As a result, the average detection efficiencies in each bin (red lines) are similar to the mean value of the whole sample (black lines, bottom rows of Figure \ref{fig:bin3_de} and \ref{fig:bin5_de}), showing that the influence caused by detection efficiencies on planet occurrence is effectively removed.

\begin{table*}
\centering
\caption{Revised characteristics at different Galactic radii ($R$) and heights ($Z$) for different Galactic components using the calibration sample updated with Gaia DR3 astrometry data.} 
{\footnotesize
\label{tab:dispersionsrevised} 
\begin{tabular}{c|c|cccccccccccc} \hline
$|Z|$ & $R$ & $\sigma^{\rm D}_{\rm U}$ & $\sigma^{\rm D}_{\rm V}$ & $\sigma^{\rm D}_{\rm W}$ & $V^{\rm D}_{\rm asym}$ & $\sigma^{\rm TD}_{\rm U}$ & $\sigma^{\rm TD}_{\rm V}$ & $\sigma^{\rm TD}_{\rm W}$ & $V^{\rm TD}_{\rm asym}$ & \multirow{2}{*}{$X_{\rm D}$} & \multirow{2}{*}{$X_{\rm TD}$} & \multirow{2}{*}{$X_{\rm H}$} & \multirow{2}{*}{$X_{\rm Herc}$}\\ 
(kpc) & (kpc) & \multicolumn{8}{c}{---------------------------~~($\rm km \ s^{-1}$)~~---------------------------}  & & & & \\ \hline 

\multirow{6}{*}{$0-0.1$}       & $7.5-8.0$  & 35 & 20 & 16 & $-14$ & 61  & 37& 31 & -41& 0.81& 0.13 & 0.0010&0.06\\
                                             & $8.0-8.5$  & 35 & 21 & 16 & $-15$ & 63  & 39 & 35 & $-44$& 0.83&0.11&0.0014&0.06\\
                                             & $8.5-9.0$  & 33 & 20 & 15 & $-13$ & 69  & 37 & 34 & $-51$& 0.85 &0.10&0.0013&0.05\\
                                             & $9.0-9.5$  & 31 & 19 & 15 & $-12$ & 69  & 35 & 33 & $-52$& 0.87&0.09&0.0012&0.04\\
                                             & $9.5-10.0$ & 28 & 19 & 15 & $-10$ & 67 & 37 & 33 & $-47$& 0.89&0.08&0.0016&0.03\\ \hline
 \multirow{6}{*}{$0.1-0.2$}    & $7.5-8.0$  & 36 & 21 & 16 & $-15$ & 63  &  37  & 36 & $-44$&0.78&0.16&0.0015&0.06\\
                                              & $8.0-8.5$  & 36  & 22 & 17 & $-16$ & 64  & 39 & 36  & -45&0.78&0.14&0.0014&0.08\\
                                              & $8.5-9.0$  & 34  & 20 & 16 & $-15$ & 70  & 38 & 36 & -53&0.82&0.12&0.0014&0.06\\
                                              & $9.0-9.5$  & 32  & 20 & 15 & $-13$ & 70  & 39 & 34 & $-52$&0.84&0.11&0.0015&0.05\\
                                              & $9.5-10.0$ & 29 & 19 & 15 & $-11$ & 69 & 37 & 34 &$-51$&0.86&0.10&0.0019&0.04\\ \hline
\multirow{6}{*}{$0.2-0.3$}    & $7.5-8.0$  & 37 & 23 & 17 & $-15$ & 66  & 39 & 37 & $-46$&0.74&0.19&0.0016&0.07\\
                                             & $8.0-8.5$  & 38 & 23 & 18 & $-16$ & 67  & 42 & 40 & $-47$&0.75&0.17&0.0015&0.08\\
                                             & $8.5-9.0$  & 36 & 21 & 17 & $-15$ & 71  & 41 & 37 &$-54$&0.79&0.15&0.0017&0.06\\
                                             & $9.0-9.5$  & 33 & 21 & 16 & $-13$ & 70  & 40 & 36 &$-52$&0.81&0.14&0.0016&0.05\\
                                             & $9.5-10.0$ & 30 & 20 & 15 & $-11$& 66  & 39 & 35 &$-47$&0.84&0.12&0.0023&0.04 \\ \hline
\multirow{6}{*}{$0.3-0.4$}     & $7.5-8.0$  & 38 & 23 & 18 & $-16$ & 67  & 41  & 38 &-48&0.70&0.23&0.0016&0.07\\
                                              & $8.0-8.5$  & 40 & 23 & 18 & $-18$ & 67  & 41 & 41 &$-51$&0.72&0.20&0.0020&0.08\\
                                              & $8.5-9.0$  & 37 & 22 & 17 & $-15$ & 70  & 41 & 38 &$-53$&0.75&0.18&0.0020&0.07\\
                                              & $9.0-9.5$  & 35 & 21 & 17 & $-14$ & 70  & 41 & 37 &$-52$&0.78&0.17&0.0020&0.05\\
                                              & $9.5-10.0$ & 34 & 20 & 16 & $-13$ & 70 & 40 & 35 &$-52$&0.80&0.15&0.0025&0.05\\ \hline
\multirow{6}{*}{$0.4-0.55$}     & $7.5-8.0$  & 42 & 25 & 19 & $-19$ & 70  & 42  & 40 &-53&0.65&0.28&0.0021&0.07\\
                                              & $8.0-8.5$  & 41 & 23 & 19 & $-18$ & 69  & 41 & 42 &$-51$&0.66&0.25&0.0026&0.09\\
                                              & $8.5-9.0$  & 39 & 23 & 19 & $-16$ & 70  & 42 & 38 &$-53$&0.70&0.23&0.0022&0.07\\
                                              & $9.0-9.5$  & 39 & 22 & 18 & $-17$ & 70  & 42 & 38 &$-53$&0.73&0.21&0.0024&0.06\\
                                              & $9.5-10.0$ & 35 & 21 & 17 & $-15$ & 69 & 40 & 35 &$-51$&0.76&0.19&0.0026&0.05\\ \hline
\multirow{6}{*}{$0.55-0.75$}     & $7.5-8.0$  & 43 & 26 & 20 & $-21$ & 70  & 45  & 41 &-52&0.55&0.36&0.0051&0.09\\
                                              & $8.0-8.5$  & 43 & 24 & 20 & $-21$  & 71  & 43 & 42 &$-54$&0.58&0.32&0.0042&0.09\\
                                              & $8.5-9.0$  & 41 & 24 & 19 & $-19$ & 71 & 43 & 40 &$-54$&0.62&0.30&0.0026&0.08\\
                                              & $9.0-9.5$  & 39 & 22 & 18 & $-17$ & 70  & 41 & 40 &$-53$&0.66&0.28&0.0026&0.06\\
                                              & $9.5-10.0$ & 38 & 21 & 17 & $-16$ & 70 & 41 & 40 &$-53$&0.68&0.25&0.0029&0.07\\ \hline
\multirow{6}{*}{$0.75-1.0$}     & $7.5-8.0$  & 44 & 28 & 23 & $-20$ & 70  & 43  & 44 &-53&0.44&0.46&0.0069&0.09\\
                                              & $8.0-8.5$  & 45 & 25 & 21 & $-22$ & 72  & 43 & 43 &$-56$&0.47&0.43&0.0078&0.09\\
                                              & $8.5-9.0$  & 41 & 25 & 20 & $-19$ & 72  & 45 & 41 &$-56$&0.50&0.41&0.0048&0.09\\
                                              & $9.0-9.5$  & 40 & 23 & 19 & $-18$ & 71  & 43 & 40 &$-54$&0.54&0.38&0.0034&0.08\\
                                              & $9.5-10.0$  & 39 & 22 & 19 & $-17$ & 72  & 42 & 40 &$-55$&0.58&0.35&0.0033&0.07\\\hline

\multirow{6}{*}{$1.0-1.5$}     & $7.5-8.0$  & 47 & 33 & 25 & $-25$ & 71  & 45  & 46 &-54&0.25&0.63&0.0169&0.10\\
                                              & $8.0-8.5$  & 46& 30 & 24 & $-22$ & 72  & 45 & 44 &$-56$&0.26&0.62&0.0157&0.10\\
                                              & $8.5-9.0$  & 43 & 29 & 22 & $-20$ & 72  & 45 & 43 &$-55$&0.32&0.58&0.0151&0.09\\
                                              & $9.0-9.5$  & 41 & 26 & 21 & $-19$ & 73  & 45 & 40 &$-57$&0.35&0.56&0.0104&0.08\\
                                              & $9.0-9.5$  & 40 & 25 & 20 & $-18$ & 71  & 42 & 40 &$-54$&0.39&0.54&0.0082&0.07\\\hline \hline

\end{tabular}}
\end{table*}

\begin{table}[!t]
\renewcommand\arraystretch{1.25}
\centering
\caption{Fitting parameters of the velocity dispersion as functions of ($R,\  Z$,  i.e., Equation \ref{eUVWRZ}) and asymmetric velocity as a function of $\sigma^2_U$ (i.e., Equation \ref{eqVasym}) derived with the calibration sample using astrometry data from Gaia DR3 and DR2.}
{\footnotesize
\label{tab:eUVWparaGaiaeDR3MOST}
\begin{tabular}{cc|cc|cc} \hline
    &    &  \multicolumn{2}{c}{---------~~Gaia DR3~~---------} &  \multicolumn{2}{c}{---------~~Gaia DR2~~---------}  \\
    &    & Thin disk & Thick disk  & Thin disk & Thick disk \\  \hline \hline
   \multirow{3}{*}{$\sigma_{U}$}  &  {$b_1$} & {$65.9^{+1.1}_{-2.9}$} & {$57.7^{+5.9}_{-4.0}$} & {$63.4^{+1.3}_{-3.2}$} & {$58.4^{+6.7}_{-4.4}$}\\ 
   &  {$b_2$} & {$-3.5^{+0.3}_{-0.1}$} & {$1.2^{+0.6}_{-0.7}$} & {$-3.2^{+0.3}_{-0.2}$} & {$1.2^{+0.6}_{-0.7}$} \\ 
   &  {$b_3$} & {$6.8^{+0.5}_{-1.3}$} & {$3.9^{+0.4}_{-0.7}$} & {$7.6^{+0.5}_{-1.3}$} & {$4.1^{+0.4}_{-0.8}$} \\ \hline
   \multirow{3}{*}{$\sigma_{V}$}  &  {$b_1$} & {$41.2^{+0.6}_{-2.1}$} & {$46.3^{+4.4}_{-3.3}$} & {$41.6^{+0.7}_{-2.3}$} & {$44.9^{+5.3}_{-2.7}$}\\ 
   & {$b_2$} & {$-2.3^{+0.3}_{-0.1}$} & {$-0.8^{+0.5}_{-0.5}$} & {$-2.3^{+0.3}_{-0.1}$} & {$5.0^{+0.2}_{-0.7}$} \\ 
   &  {$b_3$} & {$5.8^{+0.3}_{-0.5}$} & {$4.8^{+0.4}_{-1.0}$} & {$5.6^{+0.2}_{-0.9}$} & {$5.2^{+0.3}_{-1.0}$} \\ \hline
   \multirow{3}{*}{$\sigma_{W}$}  &  {$b_1$} & {$29.5^{+1.1}_{-1.2}$} & {$55.2^{+1.8}_{-2.5}$} & {$27.3^{+1.3}_{-1.5}$} & {$55.8^{+1.8}_{-3.1}$}\\ 
   & {$b_2$} & {$-1.5^{+0.1}_{-0.2}$} & {$-2.1^{+0.3}_{-0.5}$} & {$-1.2^{+0.2}_{-0.1}$} & {$-2.2^{+0.4}_{-0.2}$}\\ 
   &  {$b_3$} & {$5.0^{+0.3}_{-0.8}$} & {$5.8^{+0.5}_{-0.6}$} & {$5.0^{+0.2}_{-0.7}$} & {$6.1^{+0.4}_{-1.0}$}\\ \hline
   \multicolumn{2}{c}{{$C_0$}}  &  {$-89.7^{+1.5}_{-1.5}$}  & {$-93.0^{+2.4}_{-1.6}$}  &  {$-88.5^{+1.7}_{-1.9}$}  & {$-92.5^{+2.8}_{-2.1}$} \\\hline \hline
   
\end{tabular}}
\end{table}

\begin{table}[!t]
\renewcommand\arraystretch{1.2}
\centering
\caption{Fitting parameters of the Age-Velocity dispersion Relationship with the calibration samples using astrometry data from Gaia DR3 and DR2.}
{\footnotesize
\label{tab:AVRkbGaiaeDR3MOST}
\begin{tabular}{c|ccccc} \hline
               & \multicolumn{2}{c}{---------~~$k \ \rm (km \ s^{-1})$~~---------} & \multicolumn{2}{c}{---------~~$\beta$~~---------}    \\ 
               &  value & 1 $\sigma$ interval &  value & 1 $\sigma$ interval  \\ \hline \hline
    \multicolumn{5}{c}{Gaia DR3} \\ \hline          
    $U$  & $23.74$ & (23.47, 24.65) & $0.34$ & $(0.32,0.35)$  \\
    $V$  & $12.87$ & (12.47, 13.37) & $0.42$ & $(0.40,0.44)$  \\
    $W$  & $8.29$ & (7.92, 8.63) & $0.56$ & $(0.54,0.58)$ \\
    $V_{\rm tot}$ & $27.74$ & (27.32, 28.67) & $0.39$ & $(0.37, 0.41)$ \\ \hline \hline
    \multicolumn{5}{c}{Gaia DR2, PAST \RNum{1}} \\ \hline
    $U$  & $23.66$ & (23.07, 24.32) & $0.34$ & $(0.33,0.36)$  \\
    $V$  & $12.49$ & (12.05, 12.98) & $0.43$ & $(0.41,0.45)$  \\
    $W$  & $8.50$ & (8.09, 8.97) & $0.54$ & $(0.52,0.56)$ \\
    $V_{\rm tot}$ & $27.55$ & (26.84, 28.37) & $0.40$ & $(0.38, 0.42)$ \\ \hline \hline
 
\end{tabular}}
\end{table}

\begin{figure*}[htb!]
\centering
\includegraphics[width=.8\textwidth]{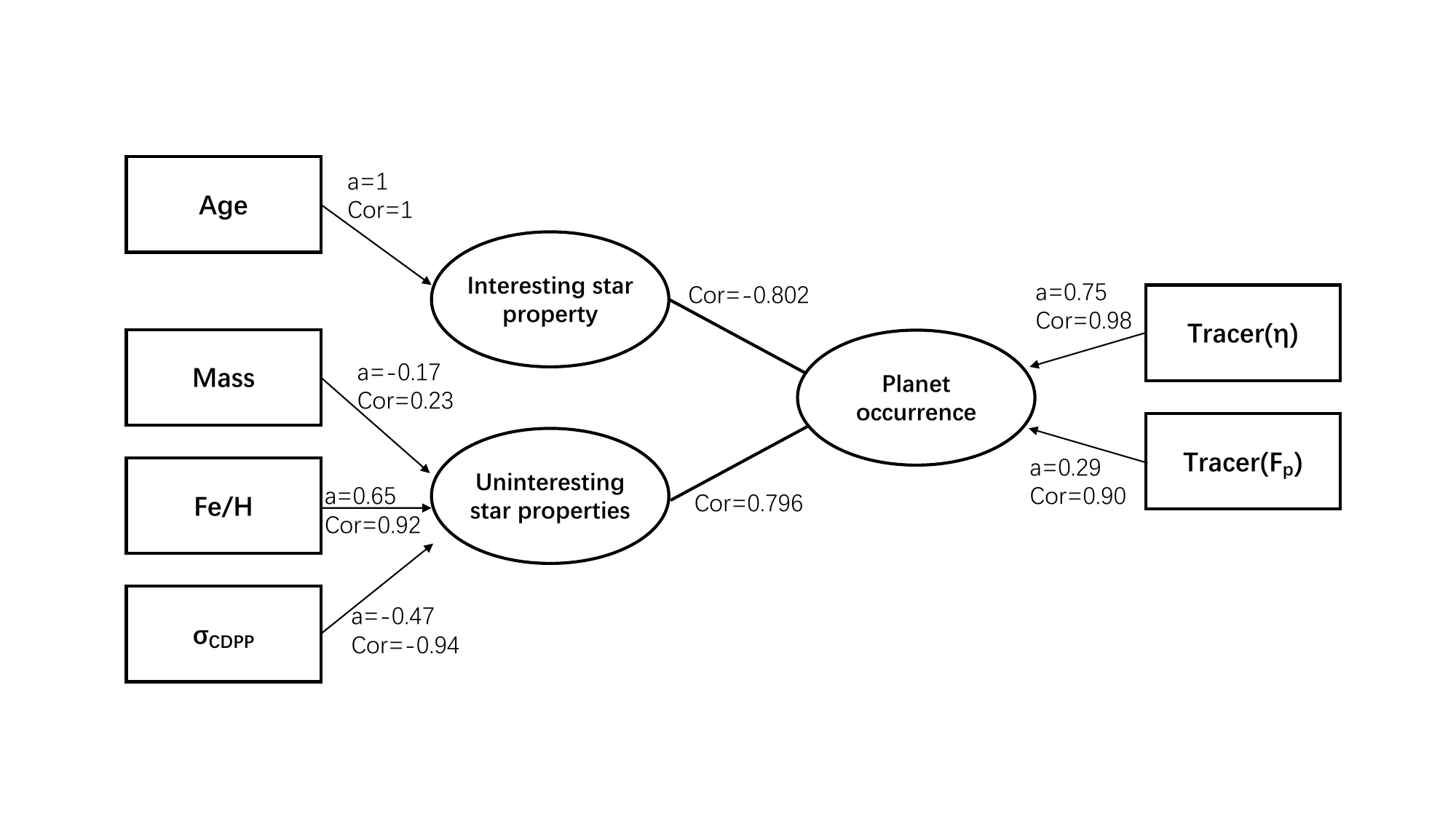}
\caption{Results of Canonical Correlation Analysis for planet and stellar properties are provided, including the weights assigned to each property and the correlations between properties and groups.
\label{fig:cca}}
\end{figure*}

\begin{figure*}[htb!]
\centering
\includegraphics[width=.6\textwidth]{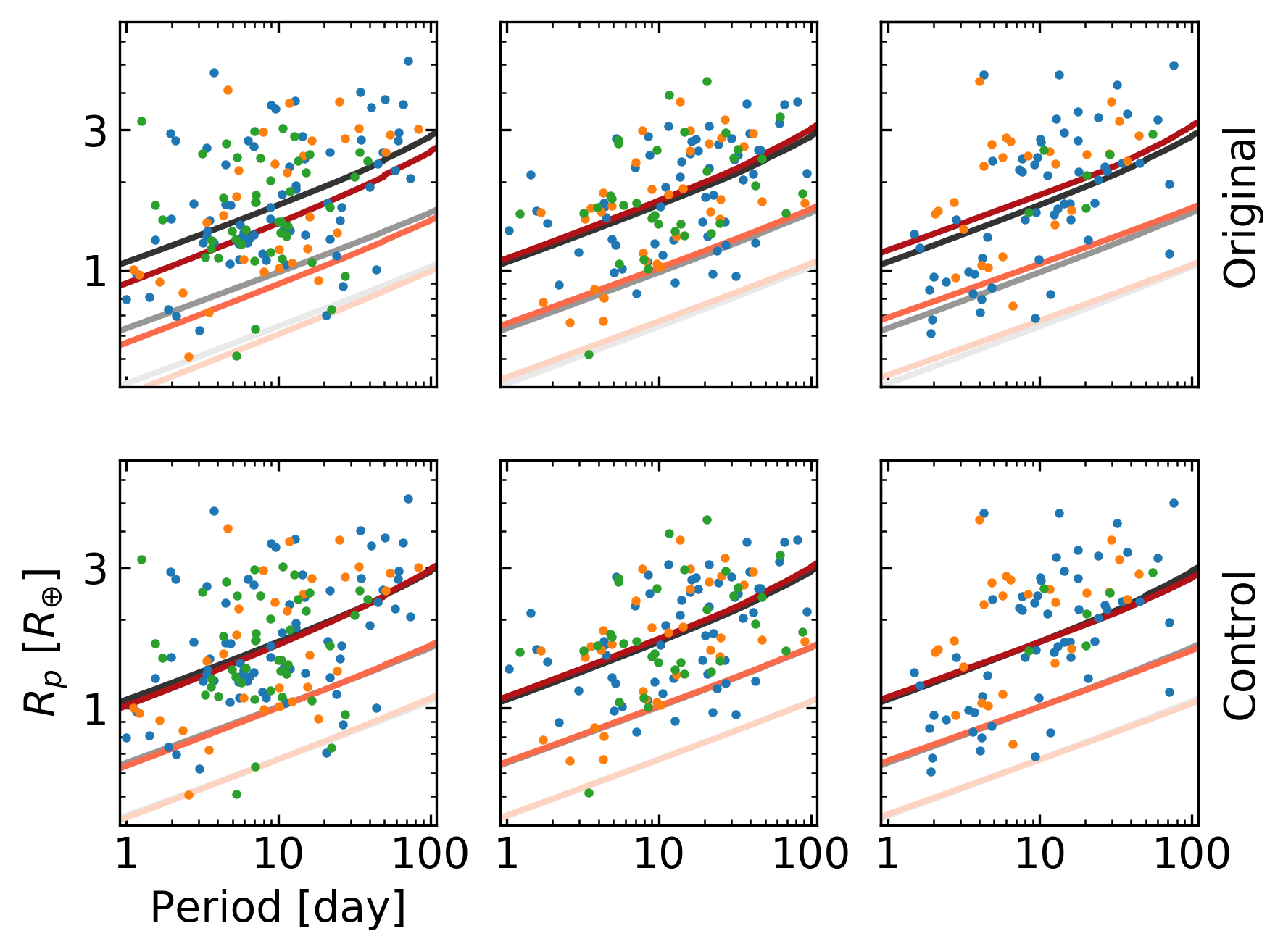}
\caption{Detection efficiencies and planet samples in the period-radius diagram for the three bins case. From top to bottom, each row corresponds to the control samples in Figure \ref{fig:bin3_cdf} and Figure \ref{fig:bin3_fs}. The reds lines present the average 90\%, 50\%, and 10\% detection efficiencies for the star sample in each bin, the grey lines show the mean 90\%, 50\%, and 10\% detection efficiencies for the whole star sample. The blue, orange, and green dots show planets in one, two, and three or more planet systems, respectively.
\label{fig:bin3_de}}
\end{figure*}

\begin{figure*}[htb!]
\centering
\includegraphics[width=.9\textwidth]{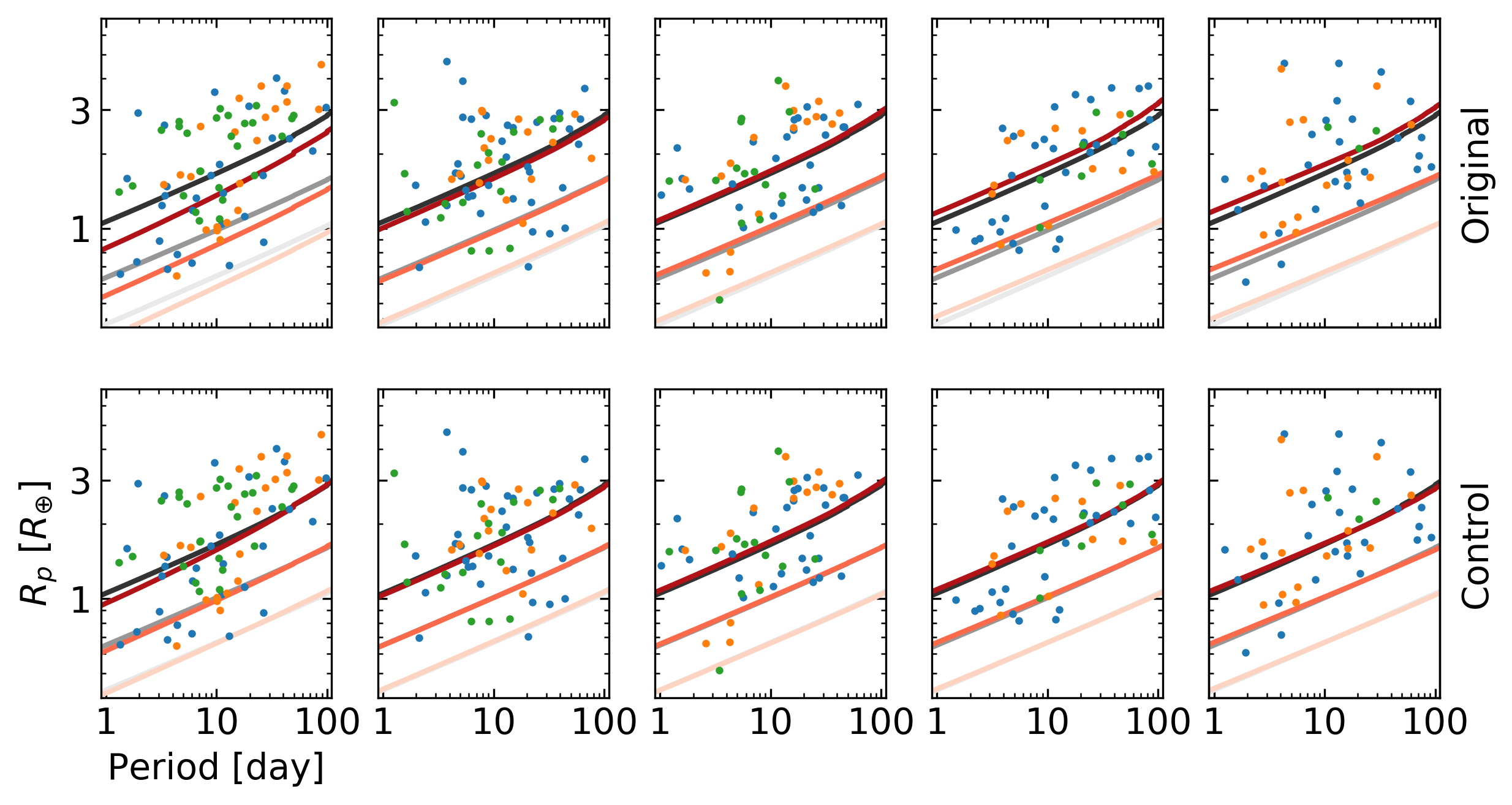}
\caption{Similar to Figure \ref{fig:bin3_de}, 90\%, 50\%, and 10\% detection efficiencies and planet samples for the five bins case. From top to bottom, each row corresponds to control samples in Figure \ref{fig:bin5_cdf} and Figure \ref{fig:bin5_fs}.
\label{fig:bin5_de}}
\end{figure*}

\clearpage
\bibliographystyle{aasjournal}
\bibliography{yjylib}

\end{document}